\documentclass{aa} 
\usepackage{natbib}
\usepackage{booktabs}
\usepackage{graphicx}
\usepackage{txfonts}
\usepackage{lipsum}
\usepackage{subcaption}         
\usepackage{lscape}             
\usepackage{placeins}           
                                
\usepackage{hyperref}

\usepackage{color}
\begin{document}

   \title{A statistical study of the environmental age of core-collapse supernovae based on VLT/MUSE integral-field-unit spectroscopy}


%

   \author{Qiang Xi\inst{1,2}
        \and Ning-Chen Sun\inst{1,2,3}\thanks{Corresponding author: sunnc@ucas.ac.cn}
        \and Yihan Zhao\inst{1,2}
        \and Emmanouil Zapartas\inst{4,5}
        \and Dimitris Souropanis\inst{5}
        \and Chun Chen \inst{6}
        \and Xiaohan Chen\inst{1,7}
        \and C\'esar Rojas-Bravo\inst{1,2}
        \and  Justyn R. Maund\inst{8}
        \and Zexi Niu\inst{1,2}
        \and Adam J. Singleton\inst{9}
        \and Anyu Wang\inst{1,2}
        \and Zhiyi Wang\inst{1,2}
        \and Ziyang Wang\inst{1,2}
        \and Junjie Wu\inst{2,1}
        \and Jifeng Liu\inst{1,2,3}
        }

   \institute{School of Astronomy and Space Science, University of Chinese Academy of Sciences, Beijing 100049, People's Republic of China
   \and National Astronomical Observatories, Chinese Academy of Sciences, Beijing 100101, People's Republic of China 
   \and Institute for Frontiers in Astronomy and Astrophysics, Beijing Normal University, Beijing, 102206, People's Republic of China
   \and Physics Department, National and Kapodistrian University of Athens, 15784 Athens, Greece
   \and Institute of Astrophysics, Foundation for Research and Technology-Hellas, GR-71110 Heraklion, Greece
   \and School of Physics and Astronomy, Sun Yat-sen University, Zhuhai 519082, China
   \and School of Physics and Astronomy, China West Normal University, Nanchong 637002, People's Republic of China
   \and Department of Physics, Royal Holloway, University of London, Egham, TW20 0EX, United Kingdom
   \and School of Mathematical and Physical Sciences, University of Sheffield, Hicks Building, Hounsfield Road, Sheffield S3 7RH, United Kingdom}

   \date{Received September 30, 20XX}

   \titlerunning{Ages of CCSNe}
   \authorrunning{Xi et al.}

 
 \abstract
{}
{We aim to understand the progenitor channels of CCSNe via a statistical study of the ages of their environments.}
{We compiled a large and minimally biased sample of 128 CCSNe discovered by untargeted wide-field transient surveys and with archival VLT/MUSE integral-field-unit spectroscopy. We measured the local H$\alpha$ luminosity within a 300-pc aperture centered on the SN explosion site as an empirical proxy for the environmental age.}
{We find that the mean local H$\alpha$ luminosities are ordered as II(P) $\approx$ IIb $\lesssim$ Ib $<$ Ic. The differences among Types~II(P), IIb and Ib are very small, if any. Type~Ic SNe are located in clearly younger environments than the other types.}
{Our result suggests that Type~Ic SNe have much younger and more massive progenitors than the other CCSN types and they likely originate from a distinct progenitor channel. The distinction between Types II(P), IIb and Ib SNe is insensitive to progenitor mass and mainly due to the different binary separation; in contrast, Type~Ic SNe predominantly require much higher-mass progenitors accompanied by close companions with large mass ratios and/or much stronger stellar wind that depends sensitively on progenitor mass.}


   \maketitle

\nolinenumbers





\section{Introduction}
\defcitealias{Xi2025a}{\emph{Paper~I}}
Core-collapse supernovae (CCSNe) mark the terminal explosions of massive stars ($M_{\rm ZAMS}\gtrsim 8~M_\odot$) and play key roles in stellar and galactic evolution. They inject $\sim 10^{51}$~erg of kinetic energy into the interstellar medium, drive turbulence and feedback, return newly synthesized heavy elements and dust to their host galaxies, and shape chemical enrichment and further star formation \citep{Smartt2009}. CCSNe also connect to multi-messenger astrophysics as potential sources of neutrinos, cosmic rays, and gravitational-wave signals from compact-remnant formation and asymmetric explosions \citep[e.g.,][]{Mirizzi2016,Helder2012,Ott2009,Abbott2020}.

Observationally, SNe are classified into Type~I and Type~II by the presence or absence of hydrogen in their spectra, respectively \citep{Minkowski1941,Filippenko1997,Turatto2003,Gal-Yam2017}. Among CCSNe, hydrogen-rich events (e.g., Type~IIP) typically arise from red supergiants that retain most of their H envelopes until core collapse \citep{Smartt2009}. In contrast, stripped-envelope (SE) SNe include Types~IIb, Ib, and Ic: Type~IIb show hydrogen only at early phases, Type~Ib lack hydrogen, and Type~Ic lack both hydrogen and conspicuous helium features. These spectroscopic differences are widely interpreted as outcomes of envelope stripping, making SESNe a uniquely sensitive probe of mass-loss processes in massive star evolution.


A key question related to SESNe is how massive stars lose their H- and He-rich envelopes before explosion. Two leading channels are commonly invoked: a single-star pathway in which line-driven winds produce massive Wolf-Rayet (WR) progenitors, and a binary pathway in which Roche-lobe overflow and/or common-envelope evolution strip the envelope over a broader range of progenitor initial masses \citep[e.g.,][]{Crowther2007,Vink2017b,Smartt2009,Yoon2015,Podsiadlowski1992,Dewi2002,Dewi2003}. Direct identification offers the most stringent constraints on SN progenitors, but for SESNe this is observationally difficult, owing to distance, crowding, extinction, and the intrinsic faintness/compactness of the stripped progenitors. The limited sample of progenitors detected so far is all consistent with the binary progenitor channel and does not provide unambiguous evidence for single WR progenitors \citep[e.g.,][]{Cao2013, Maund2004,Maund2011,Maund2019,Eldridge2015,Kilpatrick2021,Niu2024, Niu2025, Sun2020b, Sun2020a, Sun2022, Zhao2025}.

Environmental analysis provides a complementary avenue for constraining CCSN progenitors, because massive stars preferentially form in clustered environments whose stellar populations share, to first order, a common age and metallicity. Metallicity, $Z$, is particularly important because higher-metallicity stars develop stronger line-driven winds, which can more efficiently strip their outer envelopes, whereas at lower metallicity binary interaction may become the dominant stripping channel \citep{Vink2001,Eldridge2008}. Previous studies have investigated the environmental metallicities of different SN types, but no consensus has yet emerged regarding whether their metallicity distributions are systematically distinct \citep[e.g.,][]{Modjaz2011,Sanders2012,Galbany2018,Kuncarayakti2018,Pessi2023}. In \citet{Xi2025a} (hereafter \citepalias{Xi2025a}), we constructed a minimally biased, spatially resolved CCSN environmental sample from untargeted-survey discoveries with archival VLT/MUSE observations. After considering stochastic sampling effects, we found no clear type dependence in the environmental metallicity distributions.

In addition to metallicity, environmental ages have also been explored using a variety of datasets and methodologies \citep[e.g.,][]{AndersonJames2008,AndersonJames2009,Anderson2012,Williams2014,Williams2018,Galbany2018,Kuncarayakti2018,Sun2023uv,Solar2024}. Some studies have reported a monotonic progression toward younger environments with increasing envelope stripping \citep[i.e., II $>$ IIb $>$ Ib $>$ Ic; e.g.,][]{Anderson2012,Galbany2018,Kuncarayakti2018}. By contrast, other works suggest that SNe~Ic occur in significantly younger environments, while the other CCSN types exhibit broadly comparable environmental ages \citep[i.e., II $\approx$ IIb $\approx$ Ib $>$ Ic; e.g.,][]{Sun2023uv,Fang2019NatAs}. Theoretically, a recent population synthesis predicts an age sequence of II $>$ IIb $>$ Ib $>$ Ic at $Z > 0.2~Z_\odot$ while at lower metallicities the sequence changes into Ib $>$ II $>$ IIb \citep[e.g.,][]{Souropanis2025}. It is also important to note that most of the observational studies of SN environments rely on samples in the metal-rich local Universe.



Estimating ages for CCSN environments is challenging and subject to various uncertainties. Integral-field unit (IFU) spectroscopy is particularly valuable for CCSN environmental studies. Compared to traditional long-slit spectroscopy, IFU provides spatially resolved spectroscopy for the host galaxy, enabling a two-dimensional view of the complex SN environment. In practice, this allows one to identify and extract a physically motivated spectrum at (or around) the SN explosion site, to separate the nearest star-forming clump from surrounding diffuse emission, and to quantify gradients and mixing that would otherwise be folded into a single slit measurement. These capabilities are essential for robustly characterizing the immediate stellar population most plausibly linked to the progenitor.

In addition to methodology, sample-selection effects may also introduce substantial biases into the inferred demographics of CCSN environments. In order to maximize discovery rates, early SN searches often target luminous galaxies with high star formation rates. This preferentially samples younger environments and may introduce systematic biases into the resulting environmental statistics \citep{Tremonti_2004}. In contrast, modern wide-field surveys such as the Zwicky Transient Facility (ZTF; \citealp{Bellm_2019}), the Asteroid Terrestrial-impact Last Alert System (ATLAS; \citealp{Jedicke2012ALTAS}), and, in the near future, the Legacy Survey of Space and Time at the Vera C. Rubin Observatory (LSST; \citealp{Ivezic_2019}) discover transients without pre-selecting galaxies, thereby reducing this bias in SN samples.

Building on \citetalias{Xi2025a}, we use the same minimally biased CCSN sample discovered by untargeted surveys and observed with VLT/MUSE IFU spectroscopy, and extend our environmental analysis from metallicity to age. We present the environmental age distributions for different CCSN types and test whether they are statistically distinct.

This paper is organized as follows. In Section~2, we describe the methodology, including sample construction and the adopted environmental age indicators. In Section~3, we present the resulting age distributions for different CCSN types. In Section~4, we compare our findings with previous studies and discuss the implications and potential systematics. Finally, we summarize our conclusions in Section~5.

\section{Methods}

\subsection{Sample selection}

Our sample selection and data-reduction procedures follow those presented in \citetalias{Xi2025a} and we summarize the key steps here. We compiled a SN sample from the Transient Name Server\footnote{\url{https://www.wis-tns.org/}} (TNS) and the Open Supernova Catalog (OSC; \citealp{OSC2017}) and cross-matched these transients with archival VLT/MUSE observations. To mitigate potential selection biases, we imposed two principal requirements:
\begin{enumerate}
    \item \textbf{SN discovery:} To minimize host-galaxy selection effects, we restrict our sample to SNe discovered by untargeted, wide-field transient surveys (e.g., ZTF, ATLAS, or other comparable untargeted searches).
    
    \item \textbf{Data availability:} given the wide field of view of the MUSE IFU spectrograph, distant and low-mass galaxies with small angular diameters are less likely to be observed; we apply a redshift cut of $z\leq0.02$, within which the sample is quite representative of the SN population in the local Universe and the potential bias could be small (see Fig.\,2 in \citetalias{Xi2025a}).
\end{enumerate}

Starting from the sample presented in \citetalias{Xi2025a}, we further removed a number of events for which the SNe still remained bright at the epoch of the MUSE observations to avoid the contamination of SN radiation in the environmental analysis at the explosion site. After this manual vetting and the World Coordinate System (WCS)-based quality checks described below, the final sample comprises 128 SNe: 83 Type~II(P) [including SNe~IIP and Type~II events without a secure type, but most of the latter should be of Type~IIP according to the SN fraction in the local Universe; \citep{Smartt2009b,simth2011}], 14 Type~IIb, 17 Type~Ib, and 14 Type~Ic.  The vast majority of these events are located in environments with metallicities of approximately $0.2$--$1\,Z_\odot$.

\subsection{WCS correction}

Accurate astrometry is essential for an aperture-based analysis of SN environments, because the reduced MUSE datacubes can show small translational offsets between the nominal WCS and the true sky coordinates. We therefore recalibrated the astrometric zero point of the MUSE datacubes before extracting the local spectra. The correction was determined in three ways, depending on the available reference information in each field. First, when foreground stars were present in the MUSE field of view, we matched them to the Gaia DR3 or 2MASS astrometric catalogs \citep{GaiaDR3,Skrutskie2006}. Secondly, when too few suitable point sources were available, we used the galaxy nucleus or compact nuclear structure as the astrometric reference after comparison with external imaging. Thirdly, for fields with already available WCS-calibrated public MUSE products, we adopted those calibrated solutions after verifying the alignment.

In all cases, the required correction was treated as a translational offset; no rotation or pixel-scale correction was applied. We then transformed the SN coordinates from TNS/OSC onto the corrected MUSE WCS and repeated the spectral extraction and emission-line measurements at the revised explosion-site positions. This procedure ensures that the 300~pc apertures used below are centered on the best available SN locations in the MUSE cubes.

\subsection{Age indicator}

A useful age indicator is crucial for comparing CCSN environments across types. Multiple age diagnostics have been presented in the literature, including H$\alpha$ equivalent width (EW) \citep{Leloudas2011,Kuncarayakti2013_Ibc,Kuncarayakti2013_II,Kuncarayakti2018,Galbany2018,Pessi2023}, H$\alpha$-based association statistics (NCR-like measures; \citealp{AndersonJames2008,AndersonJames2009,Anderson2012}), and detailed modeling of the photoionization-sensitive emission-line ratios \citep{Sun2021,Sun2022,Sun2023}. Each of these approaches, however, has its own limitations. H$\alpha$ EW can be diluted by the continuum from older and unrelated stellar populations (see the discussion of \citealp{Sun2021}). Normalized Cumulative Rank (NCR) statistics relies on the relative rank of H$\alpha$ brightness at the SN position within the host galaxy, so it is affected by the level of star-forming activities in other parts of the galaxy. Detailed modeling of the emission lines can derive very accurate ages for the ionizing stellar population, but it is computationally expensive and can suffer from degeneracies without complementary information of the SN environment (such as high-spatial-resolution imaging of the ionizing stellar populations; e.g. see Sun et al. 2021, 2022, 2023). After a series of tests, we adopt the local H$\alpha$ luminosity, measured within a fixed physical aperture around the SN, as an indicator of progenitor age. Specifically, we adopt a physical radius of 300\,pc, motivated by the characteristic scales of clustered massive-star formation and OB associations \citep{Wright2020,Garmany1994}, the extended sizes of giant H\,II regions \citep{Kennicutt1984,Pleuss2000}, and the possibility that some progenitors may be displaced by several hundred parsecs before explosion as runaway or walkaway stars \citep{Renzo2019,Wagg2025}. Statistically, higher local H$\alpha$ luminosities on average indicate younger environments and hence higher progenitor masses, and vice versa.

We also note that the choice of a 300\,pc aperture may introduce an additional source of uncertainty. This scale should be regarded as a compromise: a smaller aperture would better preserve the locality of the measurement, but may miss emission physically associated with the progenitor because of spatial offsets, finite resolution, or progenitor migration; a larger aperture would be more inclusive, but at the cost of increased contamination from unrelated surrounding stellar populations and H$\alpha$ emission. Although we do not explicitly quantify this aperture-dependent effect here, the use of a uniform physical scale for all objects ensures that the relative comparisons across the sample remain internally consistent.

Following this idea, we processed the VLT/MUSE datacubes retrieved from the ESO Data Archive\footnote{\url{https://archive.eso.org/scienceportal/home}} with the \textsc{ifuanal} package \citep{Lyman2018}. Each cube was first corrected for foreground Milky Way extinction using the recalibration of \citet{Schlafly_2011} and the \citet{Cardelli1989} (CCM89) reddening law with $R_V=3.1$, and then shifted to the rest frame using redshifts reported by TNS/OSC. We co-added all spaxels within a circular aperture of radius 300\,pc centered on the WCS-corrected SN position. The stellar continuum was fitted and removed with the \textsc{starlight} package \citep{Fernandes2005starlight}, using the BC03 stellar population synthesis models \citep{BruzualCharlot2003}, leaving only the nebular emission lines for subsequent analysis.

\begin{figure*}
    \centering
    \includegraphics[width=1\linewidth]{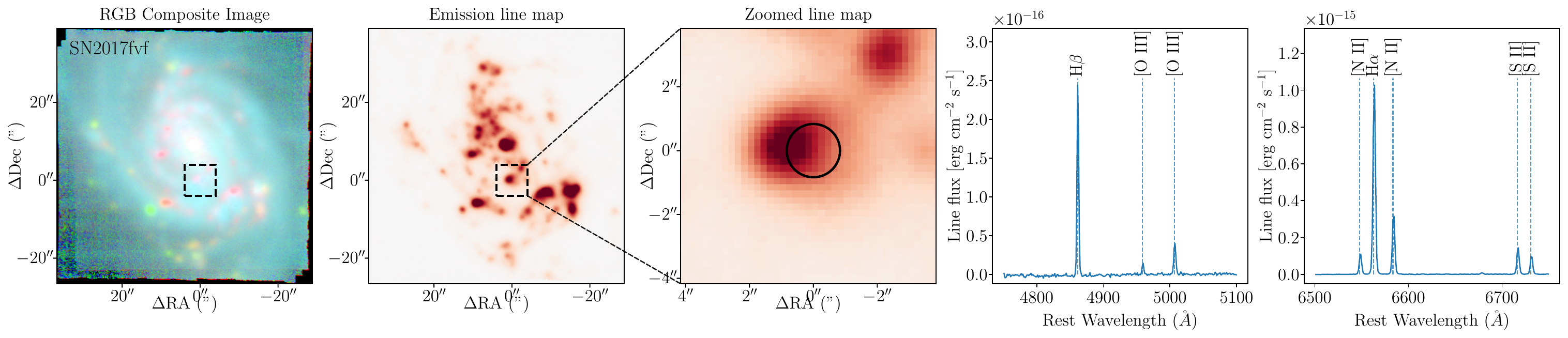}
    \includegraphics[width=1\linewidth]{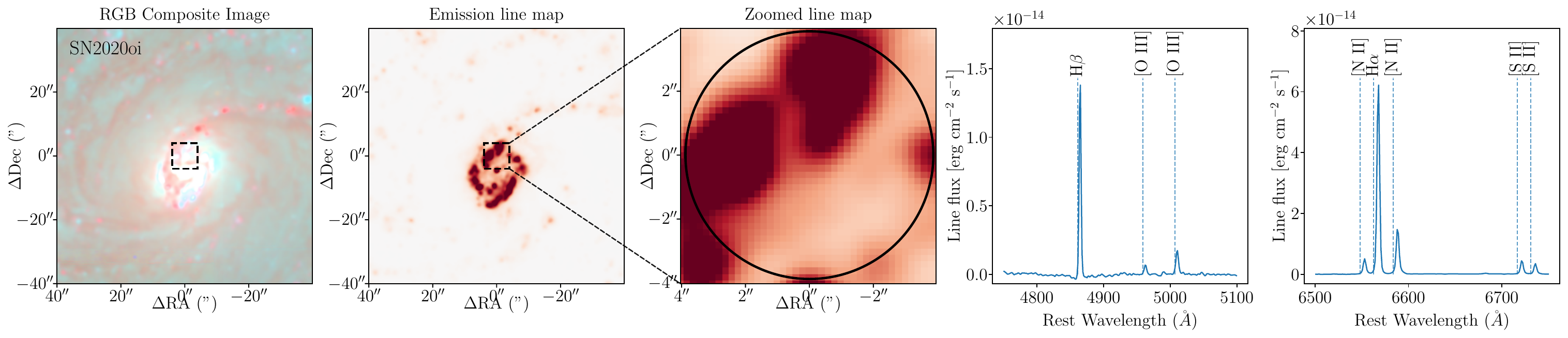}
    \includegraphics[width=1\linewidth]{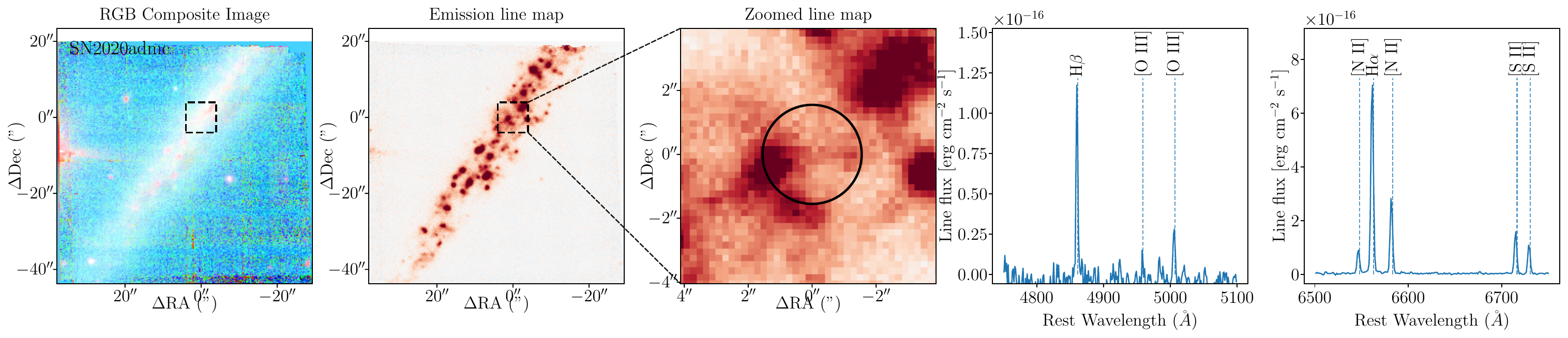}
    \caption{Examples of the MUSE-based data products used in this work. \textit{Column~1:} Pseudo-RGB composite of the host galaxy reconstructed from the MUSE datacube, where the RGB channels correspond to the cumulative fluxes in the spectral bands 6550--6750~\AA\ (R), 4950--5150~\AA\ (G), and 4750--4950~\AA\ (B). \textit{Column~2:} Continuum-subtracted nebular emission-line flux map shown on a logarithmic stretch; the color scale is in arbitrary units. \textit{Column~3:} Zoom-in on the local environment at the SN position; the black circle marks the 300~pc-radius circular aperture centered on the explosion site. \textit{Columns~4--5:} Stacked spectrum extracted within this aperture, highlighting the H$\beta$ ($\sim$4861~\AA; shown over the H$\beta$ region) and H$\alpha$ ($\sim$6563~\AA; shown over the H$\alpha$ region) complexes, respectively. The spectrum has been corrected for the host-galaxy systemic redshift and is presented in the rest frame.}
    \label{fig:example}
\end{figure*}

Fig.~\ref{fig:example} provides 3 examples of our data products. The first panel on the left shows the MUSE reconstructed pseudo-RGB composite image of the host galaxy. The second panel displays the corresponding continuum-subtracted emission-line flux map. A zoomed-in view of the SN's local environment is presented in the third panel, where the black circle indicates a 300~pc radius aperture centered on the explosion site. The final two panels on the right show the stacked spectrum extracted from this circular aperture, highlighting the H$\beta$ and H$\alpha$ emission line regions. The spectrum has been corrected for the host-galaxy systemic redshift and is presented in the rest frame.

We fitted the H$\alpha$ and H$\beta$ emission lines with single-Gaussian profiles in order to measure their integrated fluxes. The color excess due to dust in the host galaxy, $E(B\!-\!V)_{\rm host}$, was then inferred from the Balmer decrement. In the Case-B approximation, the nebula is assumed to be optically thick to Lyman-series radiation, so Lyman-line photons are absorbed and rapidly reprocessed locally, whereas Balmer and higher-series photons can escape. The canonical intrinsic ratio $(\mathrm{H}\alpha/\mathrm{H}\beta)_0 \simeq 2.86$--$2.87$ corresponds to ionized gas with an electron temperature $T_{\rm e} \simeq 10^4$~K and electron densities $n_{\rm e} \sim 10^2$--$10^4~\mathrm{cm}^{-3}$ \citep[e.g.][]{HummerStorey1987,Osterbrock2006}. Using the CCM89 reddening law, we computed
\begin{equation}
E(B\!-\!V)_{\rm host}
   = \frac{2.5}{\,k(\mathrm{H}\beta)-k(\mathrm{H}\alpha)\,}
     \log_{10}\!\left[\frac{(\mathrm{H}\alpha/\mathrm{H}\beta)_{\rm obs}}{2.87}\right],
\end{equation}
where $k(\lambda)$ denotes the value of the CCM89 reddening law at wavelength $\lambda$. We applied this color excess to deredden the spectrum and finally converted the extinction-corrected H$\alpha$ flux into the H$\alpha$ luminosity at the SN site using the host-galaxy distance. The host-galaxy distances were first taken from the HyperLEDA distance moduli, prioritizing redshift-independent estimates when available \citep{Makarov2014}. For galaxies without such measurements, we used the NASA/IPAC Extragalactic Database (NED) flow-corrected velocity $V_{\mathrm{Virgo+GA+Shapley}}$ and computed $D=V_{\mathrm{Virgo+GA+Shapley}}/H_0$, adopting $H_0=73.50\pm0.81~\mathrm{km~s^{-1}~Mpc^{-1}}$ from the H0DN Local Distance Network result \citep{Mould2000,H0DN2026}. The quoted $L_{\mathrm{H}\alpha}$ uncertainties are dominated by the formal flux-fitting errors of the H$\alpha$ and H$\beta$ emission lines, propagated through the Balmer-decrement extinction correction and the luminosity conversion.

\section{Results}

\begin{figure}
    \centering
    \includegraphics[width=1\linewidth]{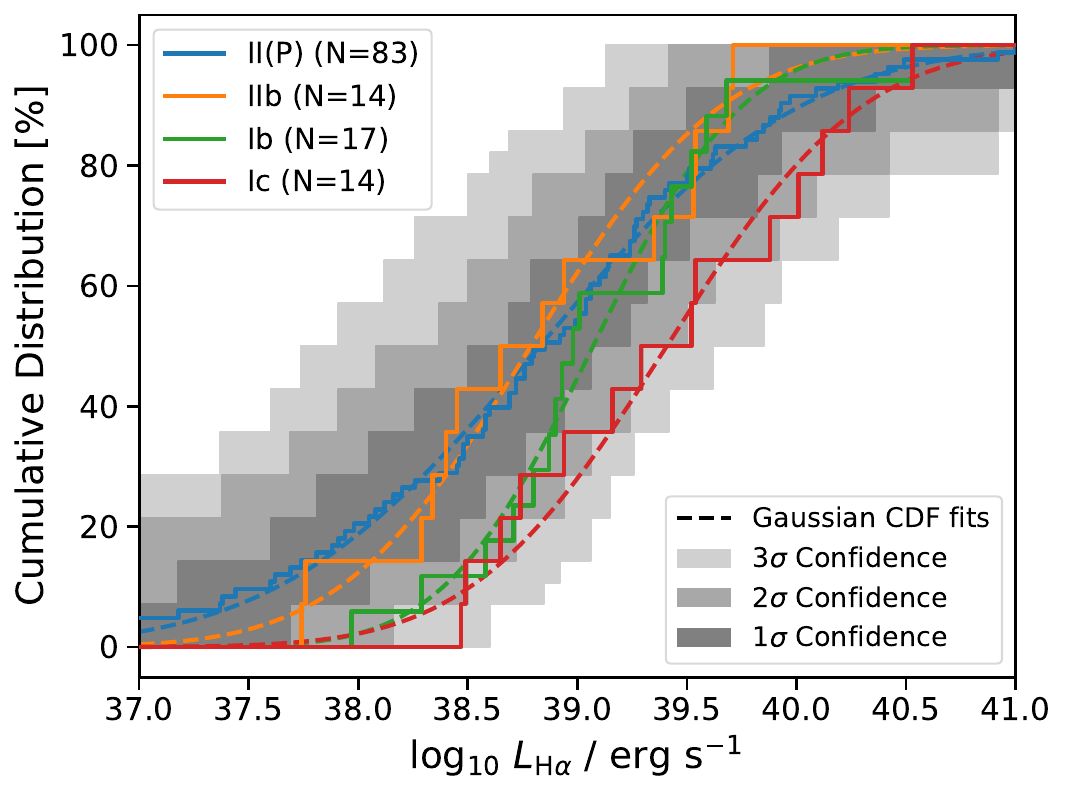}
    \caption{CDFs of the local (300~pc aperture) H$\alpha$ luminosity for different CCSN types. The blue Type~II(P) curve is used as the reference, while the grey shaded band denotes the 1$\sigma$, 2$\sigma$, and 3$\sigma$ uncertainty envelope derived from standard bootstrap resampling, with replacement, of the Type~II(P) sample at $N=14$.}
    \label{fig:CDF}
\end{figure}

Fig.~\ref{fig:CDF} shows the cumulative distributions of the local H$\alpha$ luminosity at the explosion sites of different CCSN types. The dashed curves show Gaussian fits to each type, with bootstrap resampling used to account for the measurement uncertainty of each data point during the fitting. The corresponding Gaussian-fit $\mu$ and $\sigma$ values are summarized in Tab.~\ref{tab:ha_luminosity_stats}. The detailed SN sample and the full list of measured results are available online. The local H$\alpha$ luminosities span a range of $\log_{10}(L_{\mathrm{H}\alpha}/\mathrm{erg\,s^{-1}})\approx$ 37--41. The Type~II(P) and IIb distributions have nearly identical central values, while Type~Ib is only slightly higher by $\sim 0.2$\,dex. By contrast, Type~Ic environments are systematically brighter in local H$\alpha$ luminosity by $\sim 0.5$\,dex. We also repeated this analysis after dividing the sample into several metallicity bins, and found qualitatively similar results in each bin.

To account for the stochastic sampling effect, we use Type~II(P) as the reference sample (blue curve in Fig.~\ref{fig:CDF}) and construct its confidence envelopes via the standard bootstrap resampling method. Specifically, we generated $10^{4}$ mock realizations by drawing, with replacement, $N=14$ events per realization from the Type~II(P) reference distribution; this matches the IIb and Ic sample sizes and is close to the Ib sample size of 17. The resulting $1$--$3\sigma$ confidence bands are shown as gray shading in Fig.~\ref{fig:CDF}. The Type~IIb CDF is fully consistent with the Type~II(P) reference within $1\sigma$. The Type~Ib CDF extends beyond $1\sigma$ at low H$\alpha$ luminosities but is consistent with the reference at high H$\alpha$ luminosities. The Type~Ic CDF is systematically shifted toward higher H$\alpha$ luminosities relative to the reference and extends clearly beyond the $2\sigma$ confidence envelope.

As an alternative approach, we also bootstrap--resampled the Type~II(P) reference sample while varying the draw size $N$ and, for each $N$, computed the distribution of the sample mean of local H$\alpha$ luminosity. The resulting $1\sigma$, $2\sigma$ and $3\sigma$ bands as a function of $N$ are shown in Fig.~\ref{fig:resampler}, showing that the uncertainties decrease with increasing $N$. Here $N$ is the bootstrap draw size with replacement, rather than the number of distinct objects in the Type~II(P) sample, and therefore it can exceed the number of available Type~II(P) events. Compared with Type~II(P), Type~IIb shows no difference. Type~Ib has a slightly higher mean, but the difference is not significant when the stochastic-sampling and measurement uncertainties are considered. Type~Ic is clearly higher than Type~II(P), with its mean lying beyond the $2\sigma$ stochastic-sampling band.

To assess whether the Type~Ic offset is statistically meaningful, we applied two-sample Kolmogorov--Smirnov (KS) and Anderson--Darling (AD) tests and the results are summarized as a $p$-value heatmap in Fig.~\ref{fig:AD and KS test}. Here, the $p$-value quantifies the probability of obtaining a test statistic at least as extreme as observed under the null hypothesis that the two samples are drawn from the same parent distribution. In the KS tests, we find no significant difference between any two samples.

If $p<0.05$ is adopted as the criterion, the hypothesis that Types~II(P), IIb and Ib are drawn from the same parent distribution cannot be rejected by the pairwise AD tests. In contrast, the AD test rejects the same-parent hypothesis for Type~Ic compared with II(P)+IIb+Ib, II(P)+IIb, II(P), and IIb, while the comparison between Types~Ic and Ib gives $p=0.36$. The closer agreement between Types~Ib and Ic may be related to their closer local H$\alpha$ luminosity distributions and the small subtype sample sizes.

The different results of the two tests are readily understood. The KS statistic is driven solely by the maximum pointwise separation between the empirical cumulative distribution functions (CDFs) and can have limited power when two distributions differ mainly by a broad, global shift. The AD statistic, in contrast, integrates the squared differences between the cumulative distributions with variance weighting, giving additional sensitivity to systematic offsets and tail behavior. As seen in Fig.~\ref{fig:CDF}, the Type~Ic distribution is systematically displaced to higher H$\alpha$ luminosities relative to the other types; accordingly, the AD test reports a more significant deviation than the KS test.

In summary, the central values may suggest a possible ordering of II(P) $\approx$ IIb $\lesssim$ Ib $<$ Ic in local H$\alpha$ luminosity, corresponding to II(P) $\approx$ IIb $\gtrsim$ Ib $>$ Ic in environmental age; however, the differences among Types~II(P), IIb and Ib are very small, while Type~Ic is clearly younger.
We emphasize here that the local H$\alpha$ luminosity is only an indicator instead of a precise measurement of the progenitor age. The mapping between H$\alpha$ luminosity and progenitor age is only statistically meaningful and suffers from uncertainties such as chance alignment, leakage of ionizing photons, and variations in the detailed star-formation history. As a result, the underlying difference in progenitor age and mass between Type~Ic and other SN types may be even more significant.

\begin{table}
\centering
\small
\caption{Gaussian-fit statistics of $\log_{10}\left(L_{\mathrm{H}\alpha}/\mathrm{erg\,s^{-1}}\right)$ in different types of CCSN environments.}
\label{tab:ha_luminosity_stats}
\begin{tabular}{cccc}
\hline
\hline
SN &  Number & $\mu$ & $\sigma$ \\
Type & & (dex) & (dex) \\
\hline
II(P) & 83 & 38.83 $\pm$ 0.11 & 0.93 $\pm$ 0.09 \\
IIb   & 14 & 38.79 $\pm$ 0.19 & 0.68 $\pm$ 0.11 \\
Ib    & 17 & 39.07 $\pm$ 0.14 & 0.53 $\pm$ 0.12 \\
Ic    & 14 & 39.41 $\pm$ 0.19 & 0.70 $\pm$ 0.11 \\
\hline
\hline
\end{tabular}
\end{table}

\begin{figure}
    \centering
    \includegraphics[width=1\linewidth]{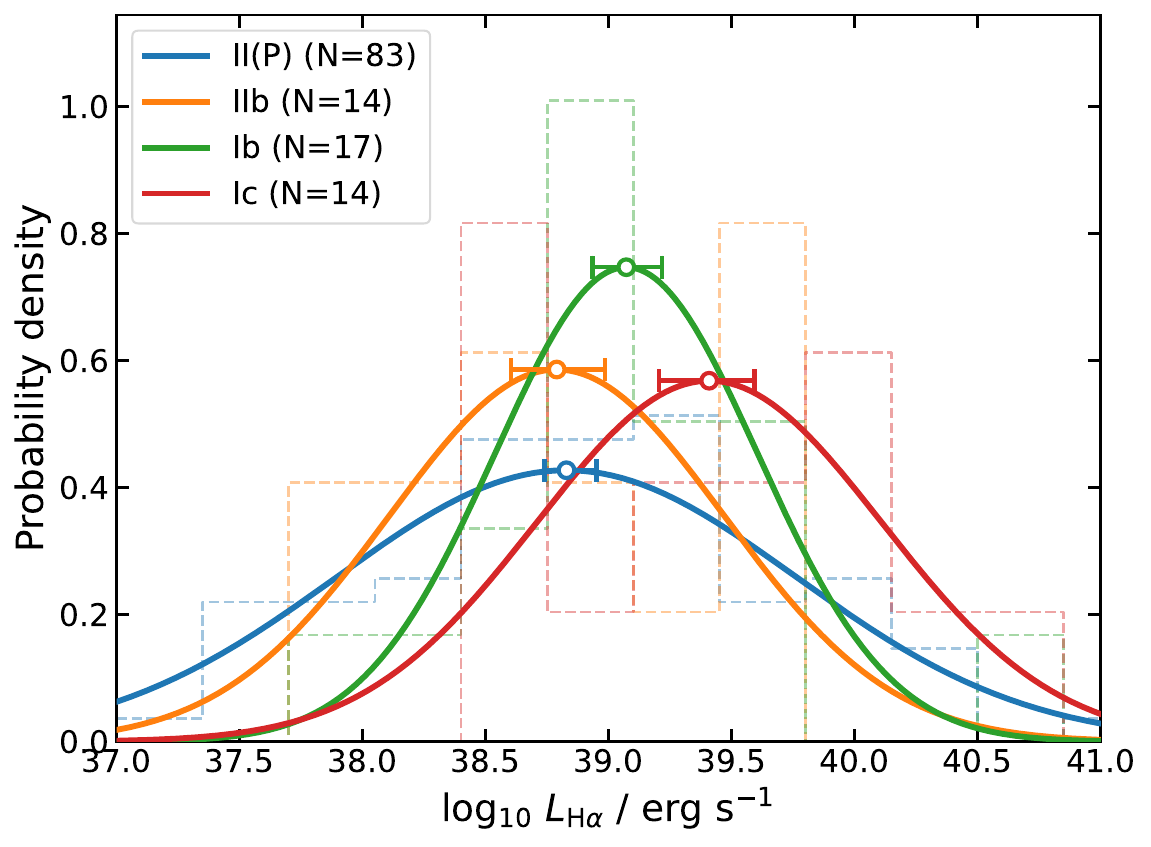}
    \caption{Histograms of the local (300~pc aperture) H$\alpha$ luminosity distributions for the different CCSN types. The histograms provide a complementary view of the CDFs in Fig.~\ref{fig:CDF}; Type~Ic SNe are shifted toward higher $\log_{10}(L_{\mathrm{H}\alpha}/\mathrm{erg\,s^{-1}})$ compared with the other types. The error bars show the Gaussian-fit mean values for each type.}
    \label{fig:histogram}
\end{figure}
\section{Discussion}
\begin{figure}
    \centering
    \includegraphics[width=1\linewidth]{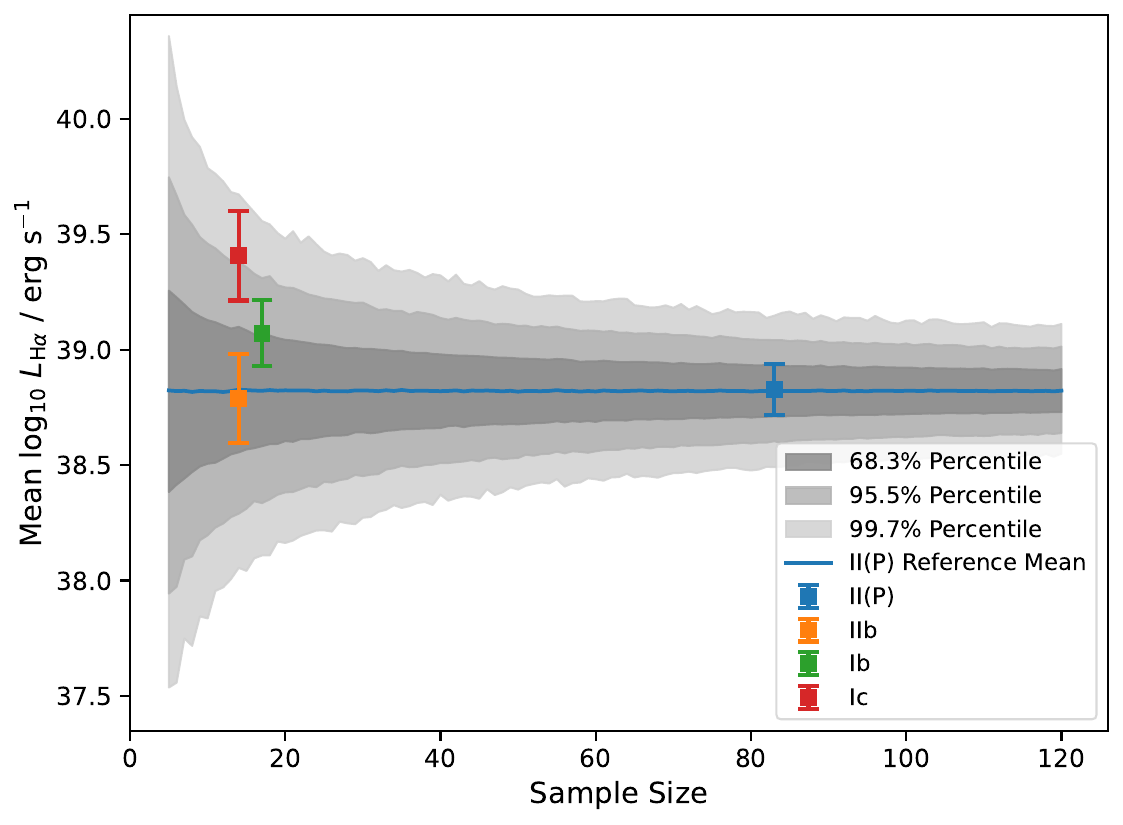}
    \caption{Mean local (300~pc aperture) H$\alpha$ luminosity as a function of bootstrap draw size. \textit{Data points:} the observed sample size and Gaussian-fit mean $\log_{10}(L_{\mathrm{H}\alpha}/\mathrm{erg\,s^{-1}})$ for each CCSN type; error bars show the bootstrap uncertainties of the Gaussian-fit means. \textit{Reference distribution:} shaded bands indicate the central 1$\sigma$, 2$\sigma$, and 3$\sigma$ intervals of the mean values obtained by randomly drawing $N$ events, with replacement, from the Type~II(P) sample. The blue curve traces the corresponding mean of the resampled distributions.}

    \label{fig:resampler}
\end{figure}

\begin{figure}
    \centering
    \includegraphics[width=1\linewidth]{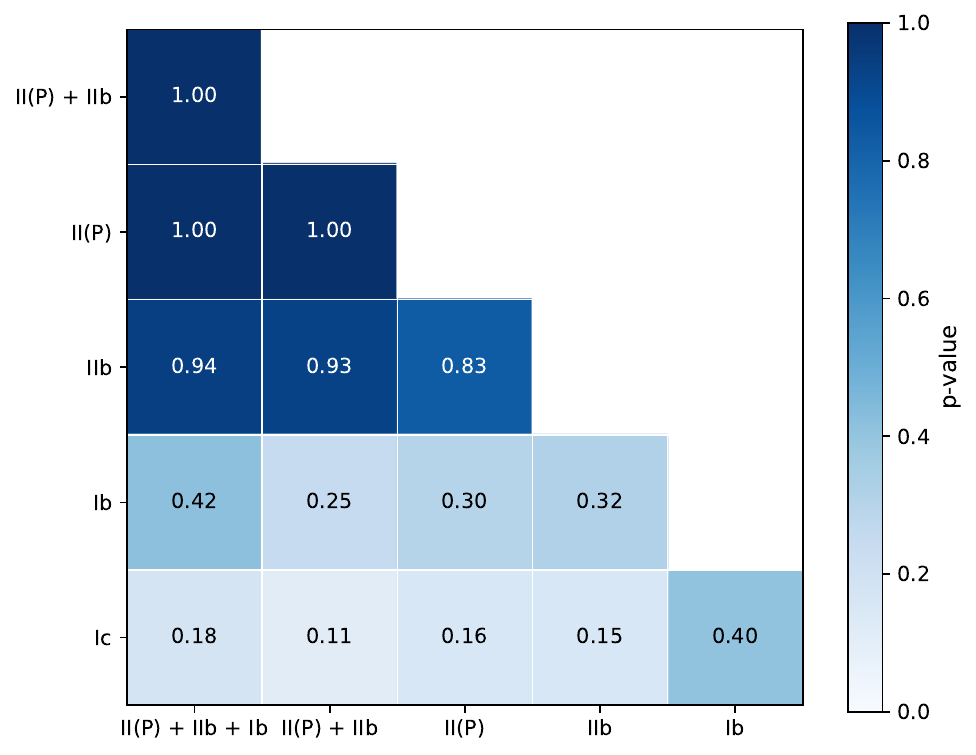}
    \includegraphics[width=1\linewidth]{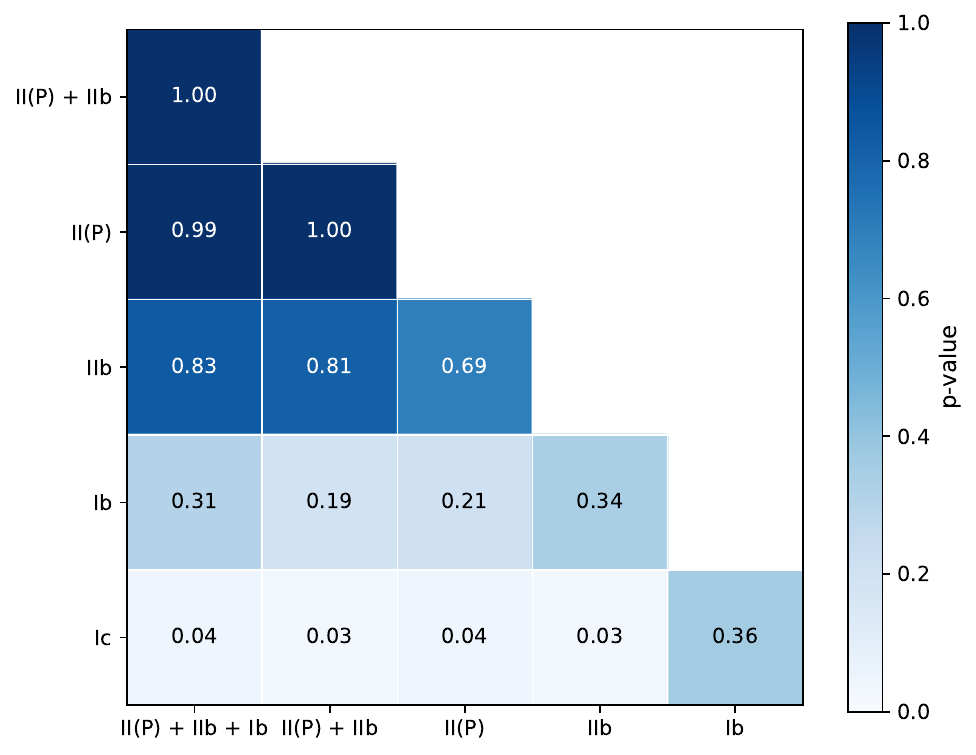}
    \caption{Heat map of pairwise $p$-values from two-sample goodness-of-fit tests comparing the CDFs of local (300~pc aperture) H$\alpha$ luminosities among CCSN types. The upper triangle reports the KS test $p$-values, while the lower triangle reports the AD test $p$-values.}
    \label{fig:AD and KS test}
\end{figure}

\subsection{Comparison with previous work}

As mentioned, CCSN environments have been analyzed in a number of previous works based on different datasets and approaches. However, the literature has not converged on a consensus on the age difference between SN types. Existing findings fall into two main categories.
Several studies argue for a monotonic sequence in which the SN environments become younger with increasing envelope stripping \citep[i.e.\ II~$>$~IIb~$>$~Ib~$>$~Ic; e.g.,][]{AndersonJames2008,Anderson2012,Kuncarayakti2018}, while other studies suggest that SNe~Ic are systematically younger than the other types and that SNe~II(P), IIb, and Ib do not have significant age differences \citep[i.e.\ II~$\approx$~IIb~$\approx$~Ib~$>$~Ic; e.g.,][]{Maund2018,Fang2019NatAs,Sun2023uv}. Our revised measurements show that the differences among Types~II(P), IIb, and Ib are very small, whereas Type~Ic is clearly younger. By using spatially resolved IFU spectroscopy, we measure the local H$\alpha$ luminosity within a uniform physical aperture, and the untargeted-discovery selection reduces host-selection biases that can affect environmental studies.

In addition, \citet{Solar2024} used molecular-gas emission as an environmental diagnostic and found no statistically significant separation among CCSN types, although CCSNe as a class occur in younger environments than SNe~Ia. Although this result may point towards the importance of binary progenitors, it may actually reflect that the particular molecular-gas tracer(s) employed are not tightly or uniquely coupled to stellar population age on the relevant timescales, thereby diluting any type-dependent age contrast.

\subsection{Implications}
Our work shows that the progenitor age and inferred initial mass are statistically indistinguishable among Types~II(P), IIb, and Ib, whereas Type~Ic SNe preferentially occur in systematically younger environments, implying higher-mass progenitors on average. This points to a distinct progenitor channel for Type~Ic compared with the bulk of CCSNe. Our sample is drawn from the local Universe ($z \leq 0.02$) and has a metallicity range of $0.2-1.0~Z_\odot$ (see \citetalias{Xi2025a}).

\citet{Souropanis2025} investigated CCSN progenitor channels with the \textsc{POSYDON} binary population-synthesis framework \citep{Fragos2023,Andrews2025}, spanning a broad range of initial masses and metallicities (see also \citealp{Claeys2011,Eldridge2017,Zapartas2017}). Under their default assumptions, they find that most Type~IIb and Type~Ib SNe originate from the initially more massive primary star in a binary (accounting for $\sim 85\%$ at solar metallicity), with a much smaller contribution from the secondary.

In contrast, most Type~II SNe arise from progenitors that are effectively single ($\sim 45.6\%$), with substantial contributions from binary mergers ($\sim 32.4\%$), consistent with the high expectations from theoretical predictions \citep{Podsiadlowski1992,Zapartas2017,Schneider2026} and with arising observational evidence of their existence \citep{Bostroem2023,Niu2026}. Only a minor fraction from the initially less massive secondary in disrupted binaries ($\sim 18.0\%$). These results imply that for moderately massive stars ($8$--$25\,M_\odot$), which dominate the CCSN rate, the SN type is comparatively less sensitive to the initial mass than to the binary separation: Type~IIb SNe are favored in wider systems where interaction leaves a small residual H envelope \citep{Claeys2011,Yoon2017,Long2022}, whereas Type~Ib SNe are favored in closer systems where the companion can remove essentially the entire H envelope \citep{Podsiadlowski1992,Dessart2024}. Stars in very wide binaries behave effectively as single stars, while those in very tight binaries often merge; both pathways predominantly yield Type~II explosions (see Fig.~1 of \citealt{Souropanis2025}). As a consequence, \citet{Souropanis2025} predict very similar progenitor ages and masses for Types~II, IIb, and Ib (their Fig.~6), consistent with our empirical finding.

Compared to stripping the H envelope, removing the more compact He-rich envelope to produce Type~Ic SNe is more demanding in binary evolution. In \citet{Souropanis2025}, this typically requires higher-mass progenitors (with more extended He envelopes) and/or extreme binary configurations (e.g., close systems with extreme mass ratios). Their synthesis predicts that $\lesssim 40\%$ of Type~Ic SNe arise from initially more massive primaries, while $\gtrsim 50\%$ originate from initially less massive secondaries. In the primary channel, progenitors are predominantly $\gtrsim 20\,M_\odot$ and undergo Case~A, Case~B, or Case~A/B mass transfer depending on the mass ratio. In the secondary channel, Type~Ic SNe commonly arise from secondaries of $\gtrsim 25\,M_\odot$ that experience common-envelope evolution with a black-hole companion (the compact remnant of the primary). Overall, their models predict that Type~Ic progenitors are systematically younger and more massive than those of Types~II, IIb, and Ib (their Fig.~6), again in agreement with our observational inference.

In our observations, the environmental age distributions of Types~Ib, IIb, and II show no obvious separation, whereas Type~Ic appears younger. This is not fully consistent with \citet{Souropanis2025}, who predict an IMF-weighted median explosion-age sequence of Type~Ic $<$ Ib $<$ IIb $<$ II at $Z>0.2~Z_\odot$. A plausible explanation is that an H$\alpha$-based age indicator is mainly sensitive to the most distinct, youngest Ic population, but is not sufficiently discriminating to robustly resolve the much smaller offsets among Types~Ib, IIb, and II. We also note that overlapping mass ranges do not imply identical IMF-weighted mass distributions: in \citet{Souropanis2025}, the IMF-weighted populations at solar metallicity still predict a modest shift in the median explosion age between Types~IIb and Ib despite their strong overlap. It therefore remains unclear whether such a small offset can be recovered with an H$\alpha$-based environmental age diagnostic, or whether it is effectively absorbed into the systematic uncertainties of the method.


It is worth noting that the type boundaries inferred from binary population synthesis can evolve with metallicity, and the corresponding conclusions may differ when considering a wider metallicity baseline. In \citet{Souropanis2025}, as metallicity decreases to $< 0.2\,Z_\odot$ (roughly the lower end probed by our sample), the separation thresholds dividing different SN types shift substantially, even though the overall parameter space producing stripped-envelope SNe (IIb and Ib) remains broadly similar. This highlights how metallicity governs both stellar winds and binary interactions, thereby shaping the effective boundaries between different outcomes (see \citealp{Souropanis2025}). A salient trend in their models is that the IIb progenitor distribution shifts toward higher masses at lower metallicity, while the Ib progenitor distribution shifts in the opposite direction. Consequently, a given progenitor mass may yield a Type~Ib SN at higher metallicity but a Type~IIb SN at lower metallicity (and vice versa), depending on how metallicity and binary interactions. Taken together, this suggests that an apparent ``progenitor-mass insensitivity'' across SESN types in multimetallicity samples could partly reflect metallicity-dependent effects washing out underlying population differences. While our environments extend down to only $\sim 0.2\,Z_\odot$, and the metallicity trends emphasized in their Fig.~3 may be modest over this interval, weaker winds than adopted in that study could render such metallicity-dependent behavior relevant even at comparatively higher metallicities.

In addition, the inferred Type~Ic progenitor channel depends sensitively on the classification criteria adopted in theoretical studies, particularly on how much helium can remain ``hidden'' in Ib/c explosions. In \citet{Souropanis2025}, the default criterion is deliberately conservative: Type~Ic progenitors must not only have low helium ejecta masses, but also low surface helium and nitrogen abundances. Under this assumption, Type~Ic SNe are produced predominantly by very massive progenitors, consistent with our inference that Ic SNe favor younger, higher-mass environments. By contrast, their alternative criterion classifies Ib/Ic events using the helium ejecta mass alone, without imposing the additional surface He/N constraints. This leads to a bimodal Type~Ic progenitor distribution, in which Ic SNe can arise both from very massive stars and from progenitors below $\sim 13\,M_\odot$, with the lower-mass channel becoming important at low metallicity.

Beyond binary interaction, stellar-wind mass loss may also contribute to producing Type~Ic SNe, but its magnitude remains highly uncertain and can alter the terminal envelope structure and hence the final SN type \citep[e.g.,][]{Woosley2019,Fullerton2006,Vink2017}. \citet{Fang2019NatAs} proposed a hybrid envelope-stripping scenario in which the H envelope is removed primarily through (largely mass-insensitive) binary interaction, while the He-rich envelope is further eroded mainly by (more mass-sensitive) winds from the post-interaction helium star (see also \citealt{Sun2023uv}). In this picture, the IIb/Ib distinction is set chiefly by binary separation, whereas the Ib/Ic distinction is driven largely by the progenitor mass through its control of helium-star wind mass loss \citep{Vink2017,DuttaKlencki2024,Yungelson2024}. This framework can also accommodate our empirical result. However, the wind mass-loss rates of helium stars remain poorly constrained observationally \citep{Pols2002,Vink2017,Moriya2022,Drout2023,gotberg2023}, and therefore the quantitative role of winds in the origin of Type~Ic SNe still awaits further investigation.

\section{Conclusions}

\label{sec:conclusions}

In this work, we investigate the age of environments of a large and minimally biased sample of CCSNe, which are discovered by untargeted wide-field transient surveys and with archival VLT/MUSE IFU spectroscopy. We use the local, extinction-corrected H$\alpha$ luminosity measured at the WCS-corrected explosion site as an age indicator to compare the environmental age distributions of Types~II(P), IIb, Ib and Ic SNe. We find that the differences in local H$\alpha$ luminosity and age among Types~II(P), IIb and Ib are very small, if any, while Type~Ic SN environments have clearly brighter local H$\alpha$ luminosities and younger ages. The Type~Ic difference cannot be fully explained by stochastic sampling effects.

Our result suggests that the progenitor age and mass are statistically indistinguishable for Types~II(P), IIb and Ib, while the Type~Ic SNe have systematically younger and more massive progenitors. This is consistent with the recent binary population synthesis carried out by \citet{Souropanis2025}, who find that, in high-metallicity environments,the distinction between Types~II, IIb and Ib is mainly due to binary separation while Type~Ic SNe originate predominantly from more massive progenitors accompanied by close companions with extreme mass ratios so that their helium envelope can be stripped by binary interaction; we emphasize that this conclusion may hold primarily at solar and sub-solar high metallicities, and may differ in substantially lower-metallicity regimes. Our result may also be explained by a hybrid envelope-stripping mechanism as proposed by \citet{Fang2019NatAs}, in which the hydrogen envelope stripping is dominated by the mass-insensitive binary interaction while the helium envelope stripping is due to the mass-sensitive stellar wind of the post-binary-interaction helium star.

\begin{acknowledgements}
This work is supported by NSFC Grants No.12303051 and No. 12261141690, the Strategic Priority Research Program of the Chinese Academy of Sciences, Grant No. XDB0550300, and by the China Manned Space Program with Grant No. CMS-CSST-2025-A14. ZXN is funded by the NSFC Grant No. 12303039. JFL acknowledges support from the NSFC through Grants No. 12588202 and from the New Cornerstone Science Foundation through the New Cornerstone Investigator Program and the XPLORER PRIZE.
\end{acknowledgements}

\section*{Data Availability}
Based on observations collected at the European Southern Observatory under ESO programme(s): 0100.B-0116(A), 0100.D-0341(A), 0101.B-0368(B), 0101.B-0706(A), 0101.D-0748(A), 0103.A-0637(A), 0103.D-0440(A), 0104.B-0404(A), 0104.D-0503(A), 094.B-0321(A), 095.B-0686(A), 095.D-0172(A), 096.B-0309(A), 096.D-0263(A), 096.D-0296(A), 097.B-0165(A), 097.B-0640(A), 097.D-0408(A), 099.D-0022(A), 104.20VC.001, 106.2104.001, 106.2155.001, 108.21ZY.008, 1100.B-0651(A), 1100.B-0651(B), 1100.B-0651(C), 1100.B-0651(D), 111.24UM.001, 111.24VQ.001, and 60.A-9301(A).

%
\bibliographystyle{bibtex/aa} 
\bibliography{bibtex/ref_revised} 

@ARTICLE{Niu2026,
       author = {{Niu}, Zexi and {Sun}, Ning-Chen and {Zapartas}, Emmanouil and {Souropanis}, Dimitris and {Cui}, Yingzhen and {Maund}, Justyn R. and {Andrews}, Jeff J. and {Briel}, Max M. and {Fraser}, Morgan and {Gossage}, Seth and {Kruckow}, Matthias U. and {Liotine}, Camille and {Liu}, Zhengwei and {Podsiadlowski}, Philipp and {Srivastava}, Philipp M. and {Teng}, Elizabeth and {Wang}, Xiaofeng and {Yang}, Yi and {Liu}, Jifeng},
        title = "{A binary merger product as the direct progenitor of a Type II-P supernova}",
      journal = {Science Bulletin},
         year = 2026,
        month = mar,
       volume = {71},
       number = {5},
        pages = {1023-1033},
          doi = {10.1016/j.scib.2026.01.036},
       adsurl = {https://ui.adsabs.harvard.edu/abs/2026SciBu..71.1023N}
}

@ARTICLE{Xi2025a,
       author = {{Xi}, Qiang and {Sun}, Ning-Chen and {Zhao}, Yi-Han and {Maund}, Justyn R. and {Niu}, Zexi and {Singleton}, Adam J. and {Liu}, Jifeng},
        title = "{A statistical study of the metallicity of core-collapse supernovae based on VLT/MUSE integral-field-unit spectroscopy}",
      journal = {\mnras},
         year = 2025,
        month = sep,
       volume = {542},
       number = {3},
        pages = {1852-1863},
          doi = {10.1093/mnras/staf1275},
archivePrefix = {arXiv},
       eprint = {2412.02667},
 primaryClass = {astro-ph.GA},
       adsurl = {https://ui.adsabs.harvard.edu/abs/2025MNRAS.542.1852X}
}

@ARTICLE{Modjaz2011,
       author = {{Modjaz}, M. and {Kewley}, L. and {Bloom}, J.~S. and {Filippenko}, A.~V. and {Perley}, D.~A. and {Silverman}, J.~M.},
        title = "{Progenitor Diagnostics for Stripped Core-collapse Supernovae: Measured Metallicities at Explosion Sites}",
      journal = {\apjl},
         year = 2011,
        month = mar,
       volume = {731},
       number = {1},
          eid = {L4},
        pages = {L4},
          doi = {10.1088/2041-8205/731/1/L4},
       adsurl = {https://ui.adsabs.harvard.edu/abs/2011ApJ...731L...4M}
}

@ARTICLE{OSC2017,
       author = {{Guillochon}, James and {Parrent}, Jerod and {Kelley}, Luke Zoltan and {Margutti}, Raffaella},
        title = "{An Open Catalog for Supernova Data}",
      journal = {\apj},
         year = 2017,
        month = jan,
       volume = {835},
       number = {1},
          eid = {64},
        pages = {64},
          doi = {10.3847/1538-4357/835/1/64},
archivePrefix = {arXiv},
       eprint = {1605.01054},
 primaryClass = {astro-ph.SR},
       adsurl = {https://ui.adsabs.harvard.edu/abs/2017ApJ...835...64G}
}

@ARTICLE{Lyman2018,
       author = {{Lyman}, J.~D. and {Taddia}, F. and {Stritzinger}, M.~D. and {Galbany}, L. and {Leloudas}, G. and {Anderson}, J.~P. and {Eldridge}, J.~J. and {James}, P.~A. and {Kr{\"u}hler}, T. and {Levan}, A.~J. and {Pignata}, G. and {Stanway}, E.~R.},
        title = "{Investigating the diversity of supernovae type Iax: a MUSE and NOT spectroscopic study of their environments}",
      journal = {\mnras},
         year = 2018,
        month = jan,
       volume = {473},
       number = {1},
        pages = {1359-1387},
          doi = {10.1093/mnras/stx2414},
archivePrefix = {arXiv},
       eprint = {1707.04270},
 primaryClass = {astro-ph.HE},
       adsurl = {https://ui.adsabs.harvard.edu/abs/2018MNRAS.473.1359L}
}

@ARTICLE{Schlafly_2011,
       author = {{Schlafly}, Edward F. and {Finkbeiner}, Douglas P.},
        title = "{Measuring Reddening with Sloan Digital Sky Survey Stellar Spectra and Recalibrating SFD}",
      journal = {\apj},
         year = 2011,
        month = aug,
       volume = {737},
       number = {2},
          eid = {103},
        pages = {103},
          doi = {10.1088/0004-637X/737/2/103},
archivePrefix = {arXiv},
       eprint = {1012.4804},
 primaryClass = {astro-ph.GA},
       adsurl = {https://ui.adsabs.harvard.edu/abs/2011ApJ...737..103S}
}

@ARTICLE{Cardelli1989,
       author = {{Cardelli}, Jason A. and {Clayton}, Geoffrey C. and {Mathis}, John S.},
        title = "{The Relationship between Infrared, Optical, and Ultraviolet Extinction}",
      journal = {\apj},
         year = 1989,
        month = oct,
       volume = {345},
        pages = {245},
          doi = {10.1086/167900},
       adsurl = {https://ui.adsabs.harvard.edu/abs/1989ApJ...345..245C}
}

@ARTICLE{Fernandes2005starlight,
       author = {{Cid Fernandes}, Roberto and {Mateus}, Ab{\'\i}lio and {Sodr{\'e}}, Laerte and {Stasi{\'n}ska}, Gra{\.z}yna and {Gomes}, Jean M.},
        title = "{Semi-empirical analysis of Sloan Digital Sky Survey galaxies - I. Spectral synthesis method}",
      journal = {\mnras},
         year = 2005,
        month = apr,
       volume = {358},
       number = {2},
        pages = {363-378},
          doi = {10.1111/j.1365-2966.2005.08752.x},
archivePrefix = {arXiv},
       eprint = {astro-ph/0412481},
 primaryClass = {astro-ph},
       adsurl = {https://ui.adsabs.harvard.edu/abs/2005MNRAS.358..363C}
}

@ARTICLE{Bellm_2019,
       author = {{Bellm}, Eric C. and {Kulkarni}, Shrinivas R. and {Graham}, Matthew J. and {Dekany}, Richard and {Smith}, Roger M. and {Riddle}, Reed and {Masci}, Frank J. and {Helou}, George and {Prince}, Thomas A. and {Adams}, Scott M. and {Barbarino}, C. and {Barlow}, Tom and {Bauer}, James and {Beck}, Ron and {Belicki}, Justin and {Biswas}, Rahul and {Blagorodnova}, Nadejda and {Bodewits}, Dennis and {Bolin}, Bryce and {Brinnel}, Valery and {Brooke}, Tim and {Bue}, Brian and {Bulla}, Mattia and {Burruss}, Rick and {Cenko}, S. Bradley and {Chang}, Chan-Kao and {Connolly}, Andrew and {Coughlin}, Michael and {Cromer}, John and {Cunningham}, Virginia and {De}, Kishalay and {Delacroix}, Alex and {Desai}, Vandana and {Duev}, Dmitry A. and {Eadie}, Gwendolyn and {Farnham}, Tony L. and {Feeney}, Michael and {Feindt}, Ulrich and {Flynn}, David and {Franckowiak}, Anna and {Frederick}, S. and {Fremling}, C. and {Gal-Yam}, Avishay and {Gezari}, Suvi and {Giomi}, Matteo and {Goldstein}, Daniel A. and {Golkhou}, V. Zach and {Goobar}, Ariel and {Groom}, Steven and {Hacopians}, Eugean and {Hale}, David and {Henning}, John and {Ho}, Anna Y.~Q. and {Hover}, David and {Howell}, Justin and {Hung}, Tiara and {Huppenkothen}, Daniela and {Imel}, David and {Ip}, Wing-Huen and {Ivezi{\'c}}, {\v{Z}}eljko and {Jackson}, Edward and {Jones}, Lynne and {Juric}, Mario and {Kasliwal}, Mansi M. and {Kaspi}, S. and {Kaye}, Stephen and {Kelley}, Michael S.~P. and {Kowalski}, Marek and {Kramer}, Emily and {Kupfer}, Thomas and {Landry}, Walter and {Laher}, Russ R. and {Lee}, Chien-De and {Lin}, Hsing Wen and {Lin}, Zhong-Yi and {Lunnan}, Ragnhild and {Giomi}, Matteo and {Mahabal}, Ashish and {Mao}, Peter and {Miller}, Adam A. and {Monkewitz}, Serge and {Murphy}, Patrick and {Ngeow}, Chow-Choong and {Nordin}, Jakob and {Nugent}, Peter and {Ofek}, Eran and {Patterson}, Maria T. and {Penprase}, Bryan and {Porter}, Michael and {Rauch}, Ludwig and {Rebbapragada}, Umaa and {Reiley}, Dan and {Rigault}, Mickael and {Rodriguez}, Hector and {van Roestel}, Jan and {Rusholme}, Ben and {van Santen}, Jakob and {Schulze}, S. and {Shupe}, David L. and {Singer}, Leo P. and {Soumagnac}, Maayane T. and {Stein}, Robert and {Surace}, Jason and {Sollerman}, Jesper and {Szkody}, Paula and {Taddia}, F. and {Terek}, Scott and {Van Sistine}, Angela and {van Velzen}, Sjoert and {Vestrand}, W. Thomas and {Walters}, Richard and {Ward}, Charlotte and {Ye}, Quan-Zhi and {Yu}, Po-Chieh and {Yan}, Lin and {Zolkower}, Jeffry},
        title = "{The Zwicky Transient Facility: System Overview, Performance, and First Results}",
      journal = {\pasp},
         year = 2019,
        month = jan,
       volume = {131},
       number = {995},
        pages = {018002},
          doi = {10.1088/1538-3873/aaecbe},
archivePrefix = {arXiv},
       eprint = {1902.01932},
 primaryClass = {astro-ph.IM},
       adsurl = {https://ui.adsabs.harvard.edu/abs/2019PASP..131a8002B}
}

@INPROCEEDINGS{Jedicke2012ALTAS,
       author = {{Jedicke}, Robert and {Tonry}, J. and {Veres}, P. and {Farnocchia}, D. and {Spoto}, F. and {Rest}, A. and {Wainscoat}, R.~J. and {Lee}, E.},
        title = "{ATLAS: Asteroid Terrestrial-impact Last Alert System}",
    booktitle = {AAS/Division for Planetary Sciences Meeting Abstracts \#44},
         year = 2012,
       series = {AAS/Division for Planetary Sciences Meeting Abstracts},
       volume = {44},
        month = oct,
          eid = {210.12},
        pages = {210.12},
       adsurl = {https://ui.adsabs.harvard.edu/abs/2012DPS....4421012J}
}

@ARTICLE{HummerStorey1987,
       author = {{Hummer}, D.~G. and {Storey}, P.~J.},
        title = "{Recombination-line intensities for hydrogenic ions - I. Case B calculations for H I and He II.}",
      journal = {\mnras},
         year = 1987,
        month = feb,
       volume = {224},
        pages = {801-820},
          doi = {10.1093/mnras/224.3.801},
       adsurl = {https://ui.adsabs.harvard.edu/abs/1987MNRAS.224..801H}
}

@BOOK{Osterbrock2006,
       author = {{Osterbrock}, Donald E. and {Ferland}, Gary J.},
        title = "{Astrophysics of gaseous nebulae and active galactic nuclei}",
         year = 2006,
    publisher = {University Science Books},
       adsurl = {https://ui.adsabs.harvard.edu/abs/2006agna.book.....O}
}

@ARTICLE{AndersonJames2008,
       author = {{Anderson}, J.~P. and {James}, P.~A.},
        title = "{Constraints on core-collapse supernova progenitors from correlations with H{\ensuremath{\alpha}} emission}",
      journal = {\mnras},
         year = 2008,
        month = nov,
       volume = {390},
       number = {4},
        pages = {1527-1538},
          doi = {10.1111/j.1365-2966.2008.13843.x},
archivePrefix = {arXiv},
       eprint = {0809.0236},
 primaryClass = {astro-ph},
       adsurl = {https://ui.adsabs.harvard.edu/abs/2008MNRAS.390.1527A}
}

@ARTICLE{AndersonJames2009,
       author = {{Anderson}, J.~P. and {James}, P.~A.},
        title = "{Comparisons of the radial distributions of core-collapse supernovae with those of young and old stellar populations}",
      journal = {\mnras},
         year = 2009,
        month = oct,
       volume = {399},
       number = {2},
        pages = {559-573},
          doi = {10.1111/j.1365-2966.2009.15324.x},
archivePrefix = {arXiv},
       eprint = {0907.0034},
 primaryClass = {astro-ph.CO},
       adsurl = {https://ui.adsabs.harvard.edu/abs/2009MNRAS.399..559A}
}

@ARTICLE{Anderson2012,
       author = {{Anderson}, J.~P. and {Habergham}, S.~M. and {James}, P.~A. and {Hamuy}, M.},
        title = "{Progenitor mass constraints for core-collapse supernovae from correlations with host galaxy star formation}",
      journal = {\mnras},
         year = 2012,
        month = aug,
       volume = {424},
       number = {2},
        pages = {1372-1391},
          doi = {10.1111/j.1365-2966.2012.21324.x},
archivePrefix = {arXiv},
       eprint = {1205.3802},
 primaryClass = {astro-ph.CO},
       adsurl = {https://ui.adsabs.harvard.edu/abs/2012MNRAS.424.1372A}
}

@ARTICLE{Kuncarayakti2013_Ibc,
       author = {{Kuncarayakti}, Hanindyo and {Doi}, Mamoru and {Aldering}, Greg and {Arimoto}, Nobuo and {Maeda}, Keiichi and {Morokuma}, Tomoki and {Pereira}, Rui and {Usuda}, Tomonori and {Hashiba}, Yasuhito},
        title = "{Integral Field Spectroscopy of Supernova Explosion Sites: Constraining the Mass and Metallicity of the Progenitors. I. Type Ib and Ic Supernovae}",
      journal = {\aj},
         year = 2013,
        month = aug,
       volume = {146},
       number = {2},
          eid = {30},
        pages = {30},
          doi = {10.1088/0004-6256/146/2/30},
archivePrefix = {arXiv},
       eprint = {1305.1105},
 primaryClass = {astro-ph.CO},
       adsurl = {https://ui.adsabs.harvard.edu/abs/2013AJ....146...30K}
}

@ARTICLE{Kuncarayakti2013_II,
       author = {{Kuncarayakti}, Hanindyo and {Doi}, Mamoru and {Aldering}, Greg and {Arimoto}, Nobuo and {Maeda}, Keiichi and {Morokuma}, Tomoki and {Pereira}, Rui and {Usuda}, Tomonori and {Hashiba}, Yasuhito},
        title = "{Integral Field Spectroscopy of Supernova Explosion Sites: Constraining the Mass and Metallicity of the Progenitors. II. Type II-P and II-L Supernovae}",
      journal = {\aj},
         year = 2013,
        month = aug,
       volume = {146},
       number = {2},
          eid = {31},
        pages = {31},
          doi = {10.1088/0004-6256/146/2/31},
archivePrefix = {arXiv},
       eprint = {1306.2106},
 primaryClass = {astro-ph.CO},
       adsurl = {https://ui.adsabs.harvard.edu/abs/2013AJ....146...31K}
}

@ARTICLE{Pessi2023,
       author = {{Pessi}, T. and {Prieto}, J.~L. and {Anderson}, J.~P. and {Galbany}, L. and {Lyman}, J.~D. and {Kochanek}, C. and {Dong}, S. and {Forster}, F. and {Gonz{\'a}lez-D{\'\i}az}, R. and {Gonzalez-Gaitan}, S. and {Guti{\'e}rrez}, C.~P. and {Holoien}, T.~W.-S. and {James}, P.~A. and {Jim{\'e}nez-Palau}, C. and {Johnston}, E.~J. and {Kuncarayakti}, H. and {Rosales-Ortega}, F. and {S{\'a}nchez}, S.~F. and {Schulze}, S. and {Shappee}, B.},
        title = "{A characterization of ASAS-SN core-collapse supernova environments with VLT+MUSE. I. Sample selection, analysis of local environments, and correlations with light curve properties}",
      journal = {\aap},
         year = 2023,
        month = sep,
       volume = {677},
          eid = {A28},
        pages = {A28},
          doi = {10.1051/0004-6361/202346512},
archivePrefix = {arXiv},
       eprint = {2306.11961},
 primaryClass = {astro-ph.SR},
       adsurl = {https://ui.adsabs.harvard.edu/abs/2023A&A...677A..28P}
}

@ARTICLE{Maund2018,
       author = {{Maund}, Justyn R.},
        title = "{The very young resolved stellar populations around stripped-envelope supernovae}",
      journal = {\mnras},
         year = 2018,
        month = may,
       volume = {476},
       number = {2},
        pages = {2629-2663},
          doi = {10.1093/mnras/sty093},
archivePrefix = {arXiv},
       eprint = {1712.07714},
 primaryClass = {astro-ph.SR},
       adsurl = {https://ui.adsabs.harvard.edu/abs/2018MNRAS.476.2629M}
}

@ARTICLE{Sun2023uv,
       author = {{Sun}, Ning-Chen and {Maund}, Justyn R. and {Crowther}, Paul A.},
        title = "{A UV census of the environments of stripped-envelope supernovae}",
      journal = {\mnras},
         year = 2023,
        month = may,
       volume = {521},
       number = {2},
        pages = {2860-2873},
          doi = {10.1093/mnras/stad690},
archivePrefix = {arXiv},
       eprint = {2209.05283},
 primaryClass = {astro-ph.SR},
       adsurl = {https://ui.adsabs.harvard.edu/abs/2023MNRAS.521.2860S}
}

@ARTICLE{Solar2024,
       author = {{Solar}, Mart{\'\i}n and {Micha{\l}owski}, Micha{\l} J. and {Nadolny}, Jakub and {Galbany}, Llu{\'\i}s and {Hjorth}, Jens and {Zapartas}, Emmanouil and {Sollerman}, Jesper and {Hunt}, Leslie and {Klose}, Sylvio and {Koprowski}, Maciej and {Le{\'s}niewska}, Aleksandra and {Ma{\l}kowski}, Micha{\l} and {Nicuesa Guelbenzu}, Ana M. and {Ryzhov}, Oleh and {Savaglio}, Sandra and {Schady}, Patricia and {Schulze}, Steve and {de Ugarte Postigo}, Antonio and {Vergani}, Susanna D. and {Watson}, Darach and {Wr{\'o}blewski}, Rados{\l}aw},
        title = "{Binary progenitor systems for Type Ic supernovae}",
      journal = {Nature Communications},
         year = 2024,
        month = dec,
       volume = {15},
       number = {1},
          eid = {7667},
        pages = {7667},
          doi = {10.1038/s41467-024-51863-z},
archivePrefix = {arXiv},
       eprint = {2409.01906},
 primaryClass = {astro-ph.SR},
       adsurl = {https://ui.adsabs.harvard.edu/abs/2024NatCo..15.7667S}
}

@ARTICLE{Fang2019NatAs,
       author = {{Fang}, Qiliang and {Maeda}, Keiichi and {Kuncarayakti}, Hanindyo and {Sun}, Fengwu and {Gal-Yam}, Avishay},
        title = "{A hybrid envelope-stripping mechanism for massive stars from supernova nebular spectroscopy}",
      journal = {Nature Astronomy},
         year = 2019,
        month = mar,
       volume = {3},
        pages = {434-439},
          doi = {10.1038/s41550-019-0710-6},
archivePrefix = {arXiv},
       eprint = {1808.04834},
 primaryClass = {astro-ph.HE},
       adsurl = {https://ui.adsabs.harvard.edu/abs/2019NatAs...3..434F}
}

@ARTICLE{Minkowski1941,
       author = {{Minkowski}, R.},
        title = "{Spectra of Supernovae}",
      journal = {\pasp},
         year = 1941,
        month = aug,
       volume = {53},
       number = {314},
        pages = {224},
          doi = {10.1086/125315},
       adsurl = {https://ui.adsabs.harvard.edu/abs/1941PASP...53..224M}
}

@ARTICLE{smartt2009,
       author = {{Smartt}, Stephen J.},
        title = "{Progenitors of Core-Collapse Supernovae}",
      journal = {\araa},
         year = 2009,
        month = sep,
       volume = {47},
       number = {1},
        pages = {63-106},
          doi = {10.1146/annurev-astro-082708-101737},
archivePrefix = {arXiv},
       eprint = {0908.0700},
 primaryClass = {astro-ph.SR},
       adsurl = {https://ui.adsabs.harvard.edu/abs/2009ARA&A..47...63S}
}

@ARTICLE{Souropanis2025,
       author = {{Souropanis}, D. and {Zapartas}, E. and {Pessi}, T. and {Briel}, M.~M. and {Renzo}, M. and {Guti{\'e}rrez}, C.~P. and {Andrews}, J.~J. and {Gossage}, S. and {Kruckow}, M.~U. and {Liotine}, C. and {Srivastava}, P.~M. and {Teng}, E.},
        title = "{The power of binaries on stripped-envelope supernovae across metallicity: uniform progenitor parameter space and persistently low ejecta masses, but subtype diversity}",
      journal = {\mnras},
         year = 2025,
        month = dec,
          doi = {10.1093/mnras/staf2163},
archivePrefix = {arXiv},
       eprint = {2508.21042},
 primaryClass = {astro-ph.SR},
       adsurl = {https://ui.adsabs.harvard.edu/abs/2025MNRAS.tmp.2036S}
}

@ARTICLE{Maund2011,
       author = {{Maund}, J.~R. and {Fraser}, M. and {Ergon}, M. and {Pastorello}, A. and {Smartt}, S.~J. and {Sollerman}, J. and {Benetti}, S. and {Botticella}, M.-T. and {Bufano}, F. and {Danziger}, I.~J. and {Kotak}, R. and {Magill}, L. and {Stephens}, A.~W. and {Valenti}, S.},
        title = "{The Yellow Supergiant Progenitor of the Type II Supernova 2011dh in M51}",
      journal = {\apjl},
         year = 2011,
        month = oct,
       volume = {739},
       number = {2},
          eid = {L37},
        pages = {L37},
          doi = {10.1088/2041-8205/739/2/L37},
archivePrefix = {arXiv},
       eprint = {1106.2565},
 primaryClass = {astro-ph.SR},
       adsurl = {https://ui.adsabs.harvard.edu/abs/2011ApJ...739L..37M}
}

@ARTICLE{Niu2024,
       author = {{Niu}, Zexi and {Sun}, Ning-Chen and {Liu}, Jifeng},
        title = "{Discovery of a Dusty Yellow Supergiant Progenitor for the Type IIb SN 2017gkk}",
      journal = {\apjl},
         year = 2024,
        month = jul,
       volume = {970},
       number = {1},
          eid = {L9},
        pages = {L9},
          doi = {10.3847/2041-8213/ad5f20},
archivePrefix = {arXiv},
       eprint = {2407.03721},
 primaryClass = {astro-ph.HE},
       adsurl = {https://ui.adsabs.harvard.edu/abs/2024ApJ...970L...9N}
}

@ARTICLE{Niu2025,
       author = {{Niu}, Zexi and {Sun}, Ning-Chen and {Maund}, Justyn R. and {Guo}, Zhen and {Li}, Wenxiong and {Sun}, Meng and {Liu}, Jifeng},
        title = "{Discovery of a Variable Yellow Supergiant Progenitor for the Type IIb SN 2024abfo}",
      journal = {\apjl},
         year = 2025,
        month = jul,
       volume = {987},
       number = {1},
          eid = {L10},
        pages = {L10},
          doi = {10.3847/2041-8213/ade4cd},
archivePrefix = {arXiv},
       eprint = {2504.20407},
 primaryClass = {astro-ph.HE},
       adsurl = {https://ui.adsabs.harvard.edu/abs/2025ApJ...987L..10N}
}

@ARTICLE{Cao2013,
       author = {{Cao}, Yi and {Kasliwal}, Mansi M. and {Arcavi}, Iair and {Horesh}, Assaf and {Hancock}, Paul and {Valenti}, Stefano and {Cenko}, S. Bradley and {Kulkarni}, S.~R. and {Gal-Yam}, Avishay and {Gorbikov}, Evgeny and {Ofek}, Eran O. and {Sand}, David and {Yaron}, Ofer and {Graham}, Melissa and {Silverman}, Jeffrey M. and {Wheeler}, J. Craig and {Marion}, G.~H. and {Walker}, Emma S. and {Mazzali}, Paolo and {Howell}, D. Andrew and {Li}, K.~L. and {Kong}, A.~K.~H. and {Bloom}, Joshua S. and {Nugent}, Peter E. and {Surace}, Jason and {Masci}, Frank and {Carpenter}, John and {Degenaar}, Nathalie and {Gelino}, Christopher R.},
        title = "{Discovery, Progenitor and Early Evolution of a Stripped Envelope Supernova iPTF13bvn}",
      journal = {\apjl},
         year = 2013,
        month = sep,
       volume = {775},
       number = {1},
          eid = {L7},
        pages = {L7},
          doi = {10.1088/2041-8205/775/1/L7},
archivePrefix = {arXiv},
       eprint = {1307.1470},
 primaryClass = {astro-ph.SR},
       adsurl = {https://ui.adsabs.harvard.edu/abs/2013ApJ...775L...7C}
}

@ARTICLE{Eldridge2015,
       author = {{Eldridge}, J.~J. and {Fraser}, Morgan and {Maund}, Justyn R. and {Smartt}, Stephen J.},
        title = "{Possible binary progenitors for the Type Ib supernova iPTF13bvn}",
      journal = {\mnras},
         year = 2015,
        month = jan,
       volume = {446},
       number = {3},
        pages = {2689-2695},
          doi = {10.1093/mnras/stu2197},
archivePrefix = {arXiv},
       eprint = {1408.4142},
 primaryClass = {astro-ph.SR},
       adsurl = {https://ui.adsabs.harvard.edu/abs/2015MNRAS.446.2689E}
}

@ARTICLE{Kilpatrick2021,
       author = {{Kilpatrick}, Charles D. and {Drout}, Maria R. and {Auchettl}, Katie and {Dimitriadis}, Georgios and {Foley}, Ryan J. and {Jones}, David O. and {DeMarchi}, Lindsay and {French}, K. Decker and {Gall}, Christa and {Hjorth}, Jens and {Jacobson-Gal{\'a}n}, Wynn V. and {Margutti}, Raffaella and {Piro}, Anthony L. and {Ramirez-Ruiz}, Enrico and {Rest}, Armin and {Rojas-Bravo}, C{\'e}sar},
        title = "{A cool and inflated progenitor candidate for the Type Ib supernova 2019yvr at 2.6 yr before explosion}",
      journal = {\mnras},
         year = 2021,
        month = jun,
       volume = {504},
       number = {2},
        pages = {2073-2093},
          doi = {10.1093/mnras/stab838},
archivePrefix = {arXiv},
       eprint = {2101.03206},
 primaryClass = {astro-ph.HE},
       adsurl = {https://ui.adsabs.harvard.edu/abs/2021MNRAS.504.2073K}
}

@ARTICLE{Sun2022,
       author = {{Sun}, Ning-Chen and {Maund}, Justyn R. and {Crowther}, Paul A. and {Hirai}, Ryosuke and {Kashapov}, Amir and {Liu}, Ji-Feng and {Liu}, Liang-Duan and {Zapartas}, Emmanouil},
        title = "{An environmental analysis of the Type Ib SN 2019yvr and the possible presence of an inflated binary companion}",
      journal = {\mnras},
         year = 2022,
        month = mar,
       volume = {510},
       number = {3},
        pages = {3701-3715},
          doi = {10.1093/mnras/stab3768},
archivePrefix = {arXiv},
       eprint = {2111.06471},
 primaryClass = {astro-ph.SR},
       adsurl = {https://ui.adsabs.harvard.edu/abs/2022MNRAS.510.3701S}
}

@ARTICLE{Zhao2025,
       author = {{Zhao}, Yi-Han and {Sun}, Ning-Chen and {Wu}, Junjie and {Niu}, Zexi and {Hong}, Xinyi and {Huang}, Yinhan and {Maund}, Justyn R. and {Xi}, Qiang and {Xiang}, Danfeng and {Liu}, Jifeng},
        title = "{Exclusion of a Direct Progenitor Detection for the Type Ic SN 2017ein Based on Late-time Observations}",
      journal = {\apjl},
         year = 2025,
        month = feb,
       volume = {980},
       number = {1},
          eid = {L6},
        pages = {L6},
          doi = {10.3847/2041-8213/adad5d},
archivePrefix = {arXiv},
       eprint = {2411.17969},
 primaryClass = {astro-ph.HE},
       adsurl = {https://ui.adsabs.harvard.edu/abs/2025ApJ...980L...6Z}
}

@ARTICLE{Maund2004,
       author = {{Maund}, Justyn R. and {Smartt}, Stephen J. and {Kudritzki}, Rolf P. and {Podsiadlowski}, Philipp and {Gilmore}, Gerard F.},
        title = "{The massive binary companion star to the progenitor of supernova 1993J}",
      journal = {\nat},
         year = 2004,
        month = jan,
       volume = {427},
       number = {6970},
        pages = {129-131},
          doi = {10.1038/nature02161},
archivePrefix = {arXiv},
       eprint = {astro-ph/0401090},
 primaryClass = {astro-ph},
       adsurl = {https://ui.adsabs.harvard.edu/abs/2004Natur.427..129M}
}

@ARTICLE{Maund2019,
       author = {{Maund}, Justyn R.},
        title = "{The Origin of the Late-time Luminosity of Supernova 2011dh}",
      journal = {\apj},
         year = 2019,
        month = sep,
       volume = {883},
       number = {1},
          eid = {86},
        pages = {86},
          doi = {10.3847/1538-4357/ab2386},
archivePrefix = {arXiv},
       eprint = {1905.08861},
 primaryClass = {astro-ph.SR},
       adsurl = {https://ui.adsabs.harvard.edu/abs/2019ApJ...883...86M}
}

@ARTICLE{Sun2020a,
       author = {{Sun}, Ning-Chen and {Maund}, Jusytn R. and {Hirai}, Ryosuke and {Crowther}, Paul A. and {Podsiadlowski}, Philipp},
        title = "{Origins of Type Ibn SNe 2006jc/2015G in interacting binaries and implications for pre-SN eruptions}",
      journal = {\mnras},
         year = 2020,
        month = feb,
       volume = {491},
       number = {4},
        pages = {6000-6019},
          doi = {10.1093/mnras/stz3431},
archivePrefix = {arXiv},
       eprint = {1909.07999},
 primaryClass = {astro-ph.SR},
       adsurl = {https://ui.adsabs.harvard.edu/abs/2020MNRAS.491.6000S}
}

@ARTICLE{Sun2020b,
       author = {{Sun}, Ning-Chen and {Maund}, Justyn R. and {Crowther}, Paul A.},
        title = "{The changing-type SN 2014C may come from an 11-M$_{{\ensuremath{\odot}}}$ star stripped by binary interaction and violent eruption}",
      journal = {\mnras},
         year = 2020,
        month = oct,
       volume = {497},
       number = {4},
        pages = {5118-5135},
          doi = {10.1093/mnras/staa2277},
archivePrefix = {arXiv},
       eprint = {2003.09325},
 primaryClass = {astro-ph.SR},
       adsurl = {https://ui.adsabs.harvard.edu/abs/2020MNRAS.497.5118S}
}

@ARTICLE{Mirizzi2016,
       author = {{Mirizzi}, A. and {Tamborra}, I. and {Janka}, H. -Th. and {Saviano}, N. and {Scholberg}, K. and {Bollig}, R. and {H{\"u}depohl}, L. and {Chakraborty}, S.},
        title = "{Supernova neutrinos: production, oscillations and detection}",
      journal = {Nuovo Cimento Rivista Serie},
         year = 2016,
        month = feb,
       volume = {39},
       number = {1-2},
        pages = {1-112},
          doi = {10.1393/ncr/i2016-10120-8},
archivePrefix = {arXiv},
       eprint = {1508.00785},
 primaryClass = {astro-ph.HE},
       adsurl = {https://ui.adsabs.harvard.edu/abs/2016NCimR..39....1M}
}

@ARTICLE{Helder2012,
       author = {{Helder}, E.~A. and {Vink}, J. and {Bykov}, A.~M. and {Ohira}, Y. and {Raymond}, J.~C. and {Terrier}, R.},
        title = "{Observational Signatures of Particle Acceleration in Supernova Remnants}",
      journal = {\ssr},
         year = 2012,
        month = nov,
       volume = {173},
       number = {1-4},
        pages = {369-431},
          doi = {10.1007/s11214-012-9919-8},
archivePrefix = {arXiv},
       eprint = {1206.1593},
 primaryClass = {astro-ph.HE},
       adsurl = {https://ui.adsabs.harvard.edu/abs/2012SSRv..173..369H}
}

@ARTICLE{Ott2009,
       author = {{Ott}, Christian D},
        title = "{TOPICAL REVIEW:  The gravitational-wave signature of core-collapse supernovae}",
      journal = {Classical and Quantum Gravity},
         year = 2009,
        month = mar,
       volume = {26},
       number = {6},
          eid = {063001},
        pages = {063001},
          doi = {10.1088/0264-9381/26/6/063001},
archivePrefix = {arXiv},
       eprint = {0809.0695},
 primaryClass = {astro-ph},
       adsurl = {https://ui.adsabs.harvard.edu/abs/2009CQGra..26f3001O}
}

@ARTICLE{Abbott2020,
       author = {{Abbott}, B.~P. and {Abbott}, R. and {Abbott}, T.~D. and {Abraham}, S. and {Acernese}, F. and {Ackley}, K. and {Adams}, C. and {Adya}, V.~B. and {Affeldt}, C. and {Agathos}, M. and {Agatsuma}, K. and {Aggarwal}, N. and {Aguiar}, O.~D. and {Aiello}, L. and {Ain}, A. and {Ajith}, P. and {Allen}, G. and {Allocca}, A. and {Aloy}, M.~A. and {Altin}, P.~A. and {Amato}, A. and {Anand}, S. and {Ananyeva}, A. and {Anderson}, S.~B. and {Anderson}, W.~G. and {Angelova}, S.~V. and {Antier}, S. and {Appert}, S. and {Arai}, K. and {Araya}, M.~C. and {Areeda}, J.~S. and {Ar{\`e}ne}, M. and {Arnaud}, N. and {Aronson}, S.~M. and {Ascenzi}, S. and {Ashton}, G. and {Aston}, S.~M. and {Astone}, P. and {Aubin}, F. and {Aufmuth}, P. and {AultONeal}, K. and {Austin}, C. and {Avendano}, V. and {Avila-Alvarez}, A. and {Babak}, S. and {Bacon}, P. and {Badaracco}, F. and {Bader}, M.~K.~M. and {Bae}, S. and {Baird}, J. and {Baker}, P.~T. and {Baldaccini}, F. and {Ballardin}, G. and {Ballmer}, S.~W. and {Bals}, A. and {Banagiri}, S. and {Barayoga}, J.~C. and {Barbieri}, C. and {Barclay}, S.~E. and {Barish}, B.~C. and {Barker}, D. and {Barkett}, K. and {Barnum}, S. and {Barone}, F. and {Barr}, B. and {Barsotti}, L. and {Barsuglia}, M. and {Barta}, D. and {Bartlett}, J. and {Bartos}, I. and {Bassiri}, R. and {Basti}, A. and {Bawaj}, M. and {Bayley}, J.~C. and {Bazzan}, M. and {B{\'e}csy}, B. and {Bejger}, M. and {Belahcene}, I. and {Bell}, A.~S. and {Beniwal}, D. and {Benjamin}, M.~G. and {Bergmann}, G. and {Bernuzzi}, S. and {Berry}, C.~P.~L. and {Bersanetti}, D. and {Bertolini}, A. and {Betzwieser}, J. and {Bhandare}, R. and {Bidler}, J. and {Biggs}, E. and {Bilenko}, I.~A. and {Bilgili}, S.~A. and {Billingsley}, G. and {Birney}, R. and {Birnholtz}, O. and {Biscans}, S. and {Bischi}, M. and {Biscoveanu}, S. and {Bisht}, A. and {Bitossi}, M. and {Bizouard}, M.~A. and {Blackburn}, J.~K. and {Blackman}, J. and {Blair}, C.~D. and {Blair}, D.~G. and {Blair}, R.~M. and {Bloemen}, S. and {Bobba}, F. and {Bode}, N. and {Boer}, M. and {Boetzel}, Y. and {Bogaert}, G. and {Bondu}, F. and {Bonnand}, R. and {Booker}, P. and {Boom}, B.~A. and {Bork}, R. and {Boschi}, V. and {Bose}, S. and {Bossilkov}, V. and {Bosveld}, J. and {Bouffanais}, Y. and {Bozzi}, A. and {Bradaschia}, C. and {Brady}, P.~R. and {Bramley}, A. and {Branchesi}, M. and {Brau}, J.~E. and {Breschi}, M. and {Briant}, T. and {Briggs}, J.~H. and {Brighenti}, F. and {Brillet}, A. and {Brinkmann}, M. and {Brockill}, P. and {Brooks}, A.~F. and {Brooks}, J. and {Brown}, D.~D. and {Brunett}, S. and {Buikema}, A. and {Bulik}, T. and {Bulten}, H.~J. and {Buonanno}, A. and {Buskulic}, D. and {Buy}, C. and {Byer}, R.~L. and {Cabero}, M. and {Cadonati}, L. and {Cagnoli}, G. and {Cahillane}, C. and {Bustillo}, J. Calder{\'o}n and {Callister}, T.~A. and {Calloni}, E. and {Camp}, J.~B. and {Campbell}, W.~A. and {Canepa}, M. and {Cannon}, K.~C. and {Cao}, H. and {Cao}, J. and {Carapella}, G. and {Carbognani}, F. and {Caride}, S. and {Carney}, M.~F. and {Carullo}, G. and {Diaz}, J. Casanueva and {Casentini}, C. and {Caudill}, S. and {Cavagli{\`a}}, M. and {Cavalier}, F. and {Cavalieri}, R. and {Cella}, G. and {Cerd{\'a}-Dur{\'a}n}, P. and {Cesarini}, E. and {Chaibi}, O. and {Chakravarti}, K. and {Chamberlin}, S.~J. and {Chan}, M. and {Chao}, S. and {Charlton}, P. and {Chase}, E.~A. and {Chassande-Mottin}, E. and {Chatterjee}, D. and {Chaturvedi}, M. and {Cheeseboro}, B.~D. and {Chen}, H.~Y. and {Chen}, X. and {Chen}, Y. and {Cheng}, H.-P. and {Cheong}, C.~K. and {Chia}, H.~Y. and {Chiadini}, F. and {Chincarini}, A. and {Chiummo}, A. and {Cho}, G. and {Cho}, H.~S. and {Cho}, M. and {Christensen}, N. and {Chu}, Q. and {Chua}, S. and {Chung}, K.~W.},
        title = "{Optically targeted search for gravitational waves emitted by core-collapse supernovae during the first and second observing runs of advanced LIGO and advanced Virgo}",
      journal = {\prd},
         year = 2020,
        month = apr,
       volume = {101},
       number = {8},
          eid = {084002},
        pages = {084002},
          doi = {10.1103/PhysRevD.101.084002},
archivePrefix = {arXiv},
       eprint = {1908.03584},
 primaryClass = {astro-ph.HE},
       adsurl = {https://ui.adsabs.harvard.edu/abs/2020PhRvD.101h4002A}
}

@ARTICLE{Crowther2007,
       author = {{Crowther}, Paul A.},
        title = "{Physical Properties of Wolf-Rayet Stars}",
      journal = {\araa},
         year = 2007,
        month = sep,
       volume = {45},
       number = {1},
        pages = {177-219},
          doi = {10.1146/annurev.astro.45.051806.110615},
archivePrefix = {arXiv},
       eprint = {astro-ph/0610356},
 primaryClass = {astro-ph},
       adsurl = {https://ui.adsabs.harvard.edu/abs/2007ARA&A..45..177C}
}

@ARTICLE{Vink2017b,
       author = {{Vink}, Jorick S.},
        title = "{Mass loss and stellar superwinds}",
      journal = {Philosophical Transactions of the Royal Society of London Series A},
         year = 2017,
        month = sep,
       volume = {375},
       number = {2105},
          eid = {20160269},
        pages = {20160269},
          doi = {10.1098/rsta.2016.0269},
archivePrefix = {arXiv},
       eprint = {1610.00578},
 primaryClass = {astro-ph.SR},
       adsurl = {https://ui.adsabs.harvard.edu/abs/2017RSPTA.37560269V}
}

@ARTICLE{Yoon2015,
       author = {{Yoon}, Sung-Chul},
        title = "{Evolutionary Models for Type Ib/c Supernova Progenitors}",
      journal = {\pasa},
         year = 2015,
        month = apr,
       volume = {32},
          eid = {e015},
        pages = {e015},
          doi = {10.1017/pasa.2015.16},
archivePrefix = {arXiv},
       eprint = {1504.01205},
 primaryClass = {astro-ph.SR},
       adsurl = {https://ui.adsabs.harvard.edu/abs/2015PASA...32...15Y}
}

@ARTICLE{Dewi2002,
       author = {{Dewi}, J.~D.~M. and {Pols}, O.~R. and {Savonije}, G.~J. and {van den Heuvel}, E.~P.~J.},
        title = "{The evolution of naked helium stars with a neutron star companion in close binary systems}",
      journal = {\mnras},
         year = 2002,
        month = apr,
       volume = {331},
       number = {4},
        pages = {1027-1040},
          doi = {10.1046/j.1365-8711.2002.05257.x},
archivePrefix = {arXiv},
       eprint = {astro-ph/0201239},
 primaryClass = {astro-ph},
       adsurl = {https://ui.adsabs.harvard.edu/abs/2002MNRAS.331.1027D}
}

@ARTICLE{Dewi2003,
       author = {{Dewi}, J.~D.~M. and {Pols}, O.~R.},
        title = "{The late stages of evolution of helium star-neutron star binaries and the formation of double neutron star systems}",
      journal = {\mnras},
         year = 2003,
        month = sep,
       volume = {344},
       number = {2},
        pages = {629-643},
          doi = {10.1046/j.1365-8711.2003.06844.x},
archivePrefix = {arXiv},
       eprint = {astro-ph/0306066},
 primaryClass = {astro-ph},
       adsurl = {https://ui.adsabs.harvard.edu/abs/2003MNRAS.344..629D}
}

@ARTICLE{Tremonti_2004,
       author = {{Tremonti}, Christy A. and {Heckman}, Timothy M. and {Kauffmann}, Guinevere and {Brinchmann}, Jarle and {Charlot}, St{\'e}phane and {White}, Simon D.~M. and {Seibert}, Mark and {Peng}, Eric W. and {Schlegel}, David J. and {Uomoto}, Alan and {Fukugita}, Masataka and {Brinkmann}, Jon},
        title = "{The Origin of the Mass-Metallicity Relation: Insights from 53,000 Star-forming Galaxies in the Sloan Digital Sky Survey}",
      journal = {\apj},
         year = 2004,
        month = oct,
       volume = {613},
       number = {2},
        pages = {898-913},
          doi = {10.1086/423264},
archivePrefix = {arXiv},
       eprint = {astro-ph/0405537},
 primaryClass = {astro-ph},
       adsurl = {https://ui.adsabs.harvard.edu/abs/2004ApJ...613..898T}
}

@ARTICLE{DuttaKlencki2024,
       author = {{Dutta}, Debasish and {Klencki}, Jakub},
        title = "{Evolutionary nature of puffed-up stripped star binaries and their occurrence in stellar populations}",
      journal = {\aap},
         year = 2024,
        month = jul,
       volume = {687},
          eid = {A215},
        pages = {A215},
          doi = {10.1051/0004-6361/202349065},
archivePrefix = {arXiv},
       eprint = {2312.12658},
 primaryClass = {astro-ph.SR},
       adsurl = {https://ui.adsabs.harvard.edu/abs/2024A&A...687A.215D}
}

@ARTICLE{Yungelson2024,
       author = {{Yungelson}, L. and {Kuranov}, A. and {Postnov}, K. and {Kuranova}, M. and {Oskinova}, L.~M. and {Hamann}, W.-R.},
        title = "{Elusive hot stripped helium stars in the Galaxy. I. Evolutionary stellar models in the gap between subdwarfs and Wolf-Rayet stars}",
      journal = {\aap},
         year = 2024,
        month = mar,
       volume = {683},
          eid = {A37},
        pages = {A37},
          doi = {10.1051/0004-6361/202347806},
       adsurl = {https://ui.adsabs.harvard.edu/abs/2024A&A...683A..37Y}
}

@ARTICLE{Ivezic_2019,
       author = {{Ivezi{\'c}}, {\v{Z}}eljko and {Kahn}, Steven M. and {Tyson}, J. Anthony and {Abel}, Bob and {Acosta}, Emily and {Allsman}, Robyn and {Alonso}, David and {AlSayyad}, Yusra and {Anderson}, Scott F. and {Andrew}, John and {Angel}, James Roger P. and {Angeli}, George Z. and {Ansari}, Reza and {Antilogus}, Pierre and {Araujo}, Constanza and {Armstrong}, Robert and {Arndt}, Kirk T. and {Astier}, Pierre and {Aubourg}, {\'E}ric and {Auza}, Nicole and {Axelrod}, Tim S. and {Bard}, Deborah J. and {Barr}, Jeff D. and {Barrau}, Aurelian and {Bartlett}, James G. and {Bauer}, Amanda E. and {Bauman}, Brian J. and {Baumont}, Sylvain and {Bechtol}, Ellen and {Bechtol}, Keith and {Becker}, Andrew C. and {Becla}, Jacek and {Beldica}, Cristina and {Bellavia}, Steve and {Bianco}, Federica B. and {Biswas}, Rahul and {Blanc}, Guillaume and {Blazek}, Jonathan and {Blandford}, Roger D. and {Bloom}, Josh S. and {Bogart}, Joanne and {Bond}, Tim W. and {Booth}, Michael T. and {Borgland}, Anders W. and {Borne}, Kirk and {Bosch}, James F. and {Boutigny}, Dominique and {Brackett}, Craig A. and {Bradshaw}, Andrew and {Brandt}, William Nielsen and {Brown}, Michael E. and {Bullock}, James S. and {Burchat}, Patricia and {Burke}, David L. and {Cagnoli}, Gianpietro and {Calabrese}, Daniel and {Callahan}, Shawn and {Callen}, Alice L. and {Carlin}, Jeffrey L. and {Carlson}, Erin L. and {Chandrasekharan}, Srinivasan and {Charles-Emerson}, Glenaver and {Chesley}, Steve and {Cheu}, Elliott C. and {Chiang}, Hsin-Fang and {Chiang}, James and {Chirino}, Carol and {Chow}, Derek and {Ciardi}, David R. and {Claver}, Charles F. and {Cohen-Tanugi}, Johann and {Cockrum}, Joseph J. and {Coles}, Rebecca and {Connolly}, Andrew J. and {Cook}, Kem H. and {Cooray}, Asantha and {Covey}, Kevin R. and {Cribbs}, Chris and {Cui}, Wei and {Cutri}, Roc and {Daly}, Philip N. and {Daniel}, Scott F. and {Daruich}, Felipe and {Daubard}, Guillaume and {Daues}, Greg and {Dawson}, William and {Delgado}, Francisco and {Dellapenna}, Alfred and {de Peyster}, Robert and {de Val-Borro}, Miguel and {Digel}, Seth W. and {Doherty}, Peter and {Dubois}, Richard and {Dubois-Felsmann}, Gregory P. and {Durech}, Josef and {Economou}, Frossie and {Eifler}, Tim and {Eracleous}, Michael and {Emmons}, Benjamin L. and {Fausti Neto}, Angelo and {Ferguson}, Henry and {Figueroa}, Enrique and {Fisher-Levine}, Merlin and {Focke}, Warren and {Foss}, Michael D. and {Frank}, James and {Freemon}, Michael D. and {Gangler}, Emmanuel and {Gawiser}, Eric and {Geary}, John C. and {Gee}, Perry and {Geha}, Marla and {Gessner}, Charles J.~B. and {Gibson}, Robert R. and {Gilmore}, D. Kirk and {Glanzman}, Thomas and {Glick}, William and {Goldina}, Tatiana and {Goldstein}, Daniel A. and {Goodenow}, Iain and {Graham}, Melissa L. and {Gressler}, William J. and {Gris}, Philippe and {Guy}, Leanne P. and {Guyonnet}, Augustin and {Haller}, Gunther and {Harris}, Ron and {Hascall}, Patrick A. and {Haupt}, Justine and {Hernandez}, Fabio and {Herrmann}, Sven and {Hileman}, Edward and {Hoblitt}, Joshua and {Hodgson}, John A. and {Hogan}, Craig and {Howard}, James D. and {Huang}, Dajun and {Huffer}, Michael E. and {Ingraham}, Patrick and {Innes}, Walter R. and {Jacoby}, Suzanne H. and {Jain}, Bhuvnesh and {Jammes}, Fabrice and {Jee}, M. James and {Jenness}, Tim and {Jernigan}, Garrett and {Jevremovi{\'c}}, Darko and {Johns}, Kenneth and {Johnson}, Anthony S. and {Johnson}, Margaret W.~G. and {Jones}, R. Lynne and {Juramy-Gilles}, Claire and {Juri{\'c}}, Mario and {Kalirai}, Jason S. and {Kallivayalil}, Nitya J. and {Kalmbach}, Bryce and {Kantor}, Jeffrey P. and {Karst}, Pierre and {Kasliwal}, Mansi M. and {Kelly}, Heather and {Kessler}, Richard and {Kinnison}, Veronica and {Kirkby}, David and {Knox}, Lloyd and {Kotov}, Ivan V. and {Krabbendam}, Victor L. and {Krughoff}, K. Simon and {Kub{\'a}nek}, Petr and {Kuczewski}, John and {Kulkarni}, Shri and {Ku}, John and {Kurita}, Nadine R. and {Lage}, Craig S. and {Lambert}, Ron and {Lange}, Travis and {Langton}, J. Brian and {Le Guillou}, Laurent and {Levine}, Deborah and {Liang}, Ming and {Lim}, Kian-Tat and {Lintott}, Chris J. and {Long}, Kevin E. and {Lopez}, Margaux and {Lotz}, Paul J. and {Lupton}, Robert H. and {Lust}, Nate B. and {MacArthur}, Lauren A. and {Mahabal}, Ashish and {Mandelbaum}, Rachel and {Markiewicz}, Thomas W. and {Marsh}, Darren S. and {Marshall}, Philip J. and {Marshall}, Stuart and {May}, Morgan and {McKercher}, Robert and {McQueen}, Michelle and {Meyers}, Joshua and {Migliore}, Myriam and {Miller}, Michelle and {Mills}, David J.},
        title = "{LSST: From Science Drivers to Reference Design and Anticipated Data Products}",
      journal = {\apj},
         year = 2019,
        month = mar,
       volume = {873},
       number = {2},
          eid = {111},
        pages = {111},
          doi = {10.3847/1538-4357/ab042c},
archivePrefix = {arXiv},
       eprint = {0805.2366},
 primaryClass = {astro-ph},
       adsurl = {https://ui.adsabs.harvard.edu/abs/2019ApJ...873..111I}
}

@ARTICLE{Leloudas2011,
       author = {{Leloudas}, G. and {Gallazzi}, A. and {Sollerman}, J. and {Stritzinger}, M.~D. and {Fynbo}, J.~P.~U. and {Hjorth}, J. and {Malesani}, D. and {Micha{\l}owski}, M.~J. and {Milvang-Jensen}, B. and {Smith}, M.},
        title = "{The properties of SN Ib/c locations}",
      journal = {\aap},
         year = 2011,
        month = jun,
       volume = {530},
          eid = {A95},
        pages = {A95},
          doi = {10.1051/0004-6361/201116692},
archivePrefix = {arXiv},
       eprint = {1102.2249},
 primaryClass = {astro-ph.SR},
       adsurl = {https://ui.adsabs.harvard.edu/abs/2011A&A...530A..95L}
}

@ARTICLE{Podsiadlowski1992,
       author = {{Podsiadlowski}, Ph. and {Joss}, P.~C. and {Hsu}, J.~J.~L.},
        title = "{Presupernova Evolution in Massive Interacting Binaries}",
      journal = {\apj},
         year = 1992,
        month = may,
       volume = {391},
        pages = {246},
          doi = {10.1086/171341},
       adsurl = {https://ui.adsabs.harvard.edu/abs/1992ApJ...391..246P}
}

\end{document}